\documentclass[12pt,a4paper]{article}

\usepackage[utf8]{inputenc}
\usepackage{authblk}
\usepackage{amsmath}
\usepackage{amsfonts}
\usepackage{amssymb}
\usepackage{graphicx}
\usepackage{hyperref}
\usepackage[left=2cm,right=2cm,top=2cm,bottom=2.5cm]{geometry}
\usepackage{bm}
\usepackage{cite}

\graphicspath{{figures/}}

\author[1]{Lauren D Smith}
\author[2]{Georg A Gottwald}
\affil[1]{Department of Mathematics, The University of Auckland, Auckland 1142, New Zealand}
\affil[2]{School of Mathematics and Statistics, The University of Sydney, Sydney, NSW 2006, Australia}
\title{Data assimilation for networks of coupled oscillators: Inferring unknown model parameters from partial observations}

\date{}

\newcommand*{\R}{{\mathbb{R}}}

\newcommand*{\bzeta}{\bm{p}}
\newcommand*{\bphi}{\bm{\phi}}

\newcommand*{\yobs}{Y}

\newcommand*{\x}{X}

\newcommand*{\X}{{\boldsymbol{X}}}

\newcommand*{\K}{K}

\newcommand*{\Pf}{P^{\rm f}}

\newcommand*{\Z}{{\boldsymbol{Z}}}

\begin{document}

\maketitle

\begin{abstract}
Inferring the state and unknown parameters of a network of coupled oscillators is of utmost importance. This task is made harder when only partial and noisy observations are available, which is a typical scenario in realistic high-dimensional systems. The general task of inference falls under data assimilation, and a commonly used assimilation method is the Ensemble Kalman Filter.
Employing network-specific localization of the forecast covariance, an Ensemble Kalman Filter with state space augmentation is shown to yield highly accurate estimates of both the oscillator phases and unknown model parameters in the case where only a subset of oscillator phases are observed. In contrast, standard data assimilation methods yield poor results.
We demonstrate the effectiveness of our approach for Kuramoto oscillators and for networks of theta neurons, using a variety of network topologies.

This is the accepted version to appear in \textit{Proceedings of the Royal Society A}. The published version will be available at doi: \href{https://doi.org/10.1098/rspa.2024.0813}{10.1098/rspa.2024.0813}
\end{abstract}



\section{Introduction} \label{sec:intro}

Many natural phenomena and engineering applications can be described as networks of coupled oscillators, for example, the firing of neurons in the brain \cite{BickEtAl2020, MontbrioEtAl2015, SchmidtAvitabile2020, GersterEtAl20} and the dynamics of power grids \cite{MachowskiEtAl2011, Eremia2013, NishikawaMotter2015, FilatrellaEtAl08, SchaferYalcin2019}. However, mathematical models for these systems are often incomplete, with unknown parameters and simplifying physical assumptions. Scientists are now seeking data-driven methods to improve estimates of the dynamic state of these systems using observational data, and to simultaneously estimate unknown parameters. 

A pertinent question is: Can one accurately estimate all the oscillators' phases as well as unknown model parameters if only a subset of the oscillator phases are observed? For example, in a power grid, we may be able to observe all the power stations, but none of the consumers. The framework of data assimilation (DA) provides a unifying framework that estimates a  system's state and unknown parameters by combining two uncertain pieces of information -- the output of our imperfect model (the model forecast)  and noisy partial observations of the system at discrete time intervals. 


We focus specifically on the Ensemble Kalman Filter (EnKF)  \cite{Evensen1994, Houtekamer1998, Jazwinski2007, Evensen2009a, Evensen2009, Reich2015, LawEtAl2015} as a DA method.
Although the framework of Kalman filters can strictly only be applied to linear systems \cite{LeGlandEtAl11}, the EnKF has been successfully used in numerous non-linear settings \cite{CarrilloEtAl2024}, in particular, weather forecasting \cite{Evensen1994, Houtekamer1998, Jazwinski2007, Evensen2009a, Evensen2009, Reich2015, LawEtAl2015}.
To date, there have been only a few applications of DA to the broad class of non-linear systems that model coupled oscillators on complex networks \cite{Forero-OrtizEtAl2021, AristidesEtAl2023}. These applications, however, assume that all nodes are observed. 
 Indeed, we find that the standard EnKF performs poorly for networks of coupled oscillators unless all nodes are observed or very large ensembles are used. 
 Covariance localization, whereby potentially spurious entries of the covariance matrix are set to zero, has been recognized as an essential component for successful performance of EnKFs \cite{Houtekamer2001, SanzAlonso23}.
 In order to employ localization, one needs to identify those entries of the sample covariance matrix which are overwhelmed by finite-sample noise \cite{Houtekamer1998, MenetrierEtAl15, SnyderHakim22, VishnyEtAl24, MorzfeldEtAl19}. If two variables are not correlated or are very weakly correlated, the corresponding entry of the sample covariance matrix should be small, but is easily contaminated by finite-sample noise. One often employs a notion of distance to identify which variables are weakly correlated: the larger the distance the less their mutual effect on each other. 
 In applications involving partial differential equations on spatial domains, for example in weather forecasting, the Euclidean distance is appropriate and localization has shown to be highly efficient in mitigating detrimental small ensemble size effects. It is less clear how best to incorporate ``distance'' and localization for dynamics on networks.
 We introduce a novel localization method, specific to dynamics on networks, which renders the EnKF highly effective at estimating both oscillator phases and unknown model parameters, even when a small ensemble is used and only a fraction of the phases are observed. 
 This sets DA apart from other data-driven parameter estimation methods that require all oscillator phases to be observed \cite{PanaggioEtAl2019, ShandilyaTimme2011, Pikovsky2018, OwensKutz2022}. We demonstrate our method for the Kuramoto model \cite{Kuramoto84, Strogatz00, PikovskyEtAl01, AcebronEtAl05, OsipovEtAl07, ArenasEtAl08, DorflerBullo14, RodriguesEtAl16} which captures the synchronization phenomena observed in many coupled oscillator systems, and for networks of theta neurons  \cite{Laing2014, Laing2018, LaingOmelchenko2020, OmelchenkoLaing2022} which model neural dynamics of the brain.
 
The article is organized as follows. In Section~\ref{sec:models} we introduce the Kuramoto model and the theta neuron model, which are the models we use to test the efficacy of our DA methods. In Section~\ref{sec:EnKF} we give details of the Ensemble Kalman Filter and our novel network-specific localization procedure. In Section~\ref{sec:results} we apply our DA methods to the Kuramoto model (Section~\ref{sec:DA_KM}) and to the theta neuron model (Section~\ref{sec:DA_theta}), showing that the EnKF with network specific localization provides a great improvement over the standard EnKF. In Section~\ref{sec:conclusions} we summarize our results.

\section{Models and their dynamics} \label{sec:models}

We consider two prototypical examples of coupled phase oscillators; the Kuramoto model and networks of theta neurons. Here we describe the details of the models and their dynamics.

\subsection{The Kuramoto model}

Due to its ability to describe many real-world synchronization phenomena and its analytical tractability, the dynamics of the Kuramoto model \cite{Kuramoto84} has been widely studied \cite{Strogatz00, PikovskyEtAl01, AcebronEtAl05, OsipovEtAl07, ArenasEtAl08, DorflerBullo14, RodriguesEtAl16}. For $N$ oscillators, the dynamics of the $i$-th oscillator is
\begin{equation} \label{eq:Kuramoto}
\dot{\phi}_i = \omega_i + \frac{\kappa}{N}\sum_{j=1}^N A_{ij} \sin(\phi_j - \phi_i ) ,
\end{equation} 
where $\omega_i$ is the oscillator's natural frequency, $\kappa$ is the coupling strength and $A$ is the network adjacency matrix, such that $A_{ij}=1$ if nodes $i$ and $j$ are connected and $A_{ij}=0$ otherwise.

\begin{figure}[tbp]
\centering
\includegraphics[width=0.7\columnwidth]{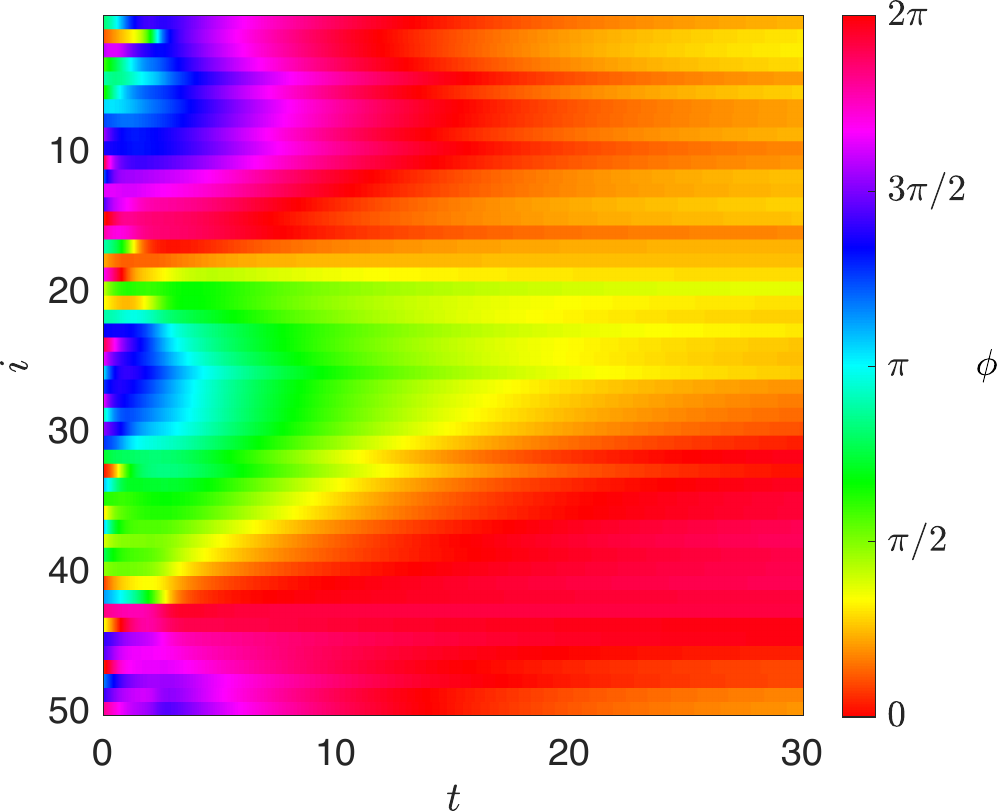}
\caption{
Time evolution of the $N=50$ phases $\phi_i$ of the Kuramoto model (\ref{eq:Kuramoto}) showing convergence to a synchronized state. The Kuramoto model parameters are $\kappa = 27$ and $\omega_i \sim \mathcal{N}(0,0.1)$, with a ring network topology $A$ with coupling radius $r=3$, where each node is connected to its $2r$ nearest neighbors.
}
\label{fig:KM_phi_of_t}
\end{figure}

By moving to a rotating reference frame, we may assume without loss of generality that the mean of the natural frequencies is zero, and in the rotating reference frame a synchronized state appears as a stationary state with all phases approximately equal. An example of this is shown in Fig.~\ref{fig:KM_phi_of_t}, such that initially random phases rapidly align and synchronize. In this case a ring network topology is used, but the same synchronization phenomenon is universal to all network topologies for sufficiently large coupling strengths $\kappa$.


We will consider in Section~\ref{sec:DA_KM} the challenge to accurately estimate all the phases $\phi_i$ and all the natural frequencies $\omega_i$, given only noisy observations of a subset of the phases and none of the frequencies.

The Kuramoto model (\ref{eq:Kuramoto}) can be thought of as an inertia-less simplification of the dynamics of power grids, wherein the oscillators of the network are the power stations, consumers and buses, and power lines form the network edges. The dynamics of power grids is described by the swing equation which can be written as the second order Kuramoto model \cite{NishikawaMotter2015, FilatrellaEtAl08, SchaferYalcin2019} which includes inertia. Real-time observational data can be obtained from the nodes in the power grid, such as their production or consumption, their operating frequency, and their phase difference from the local net phase \cite{MachowskiEtAl2011, Eremia2013}.

\subsection{Networks of theta neurons}

\begin{figure}[tbp]
\centering
\includegraphics[width=0.7\columnwidth]{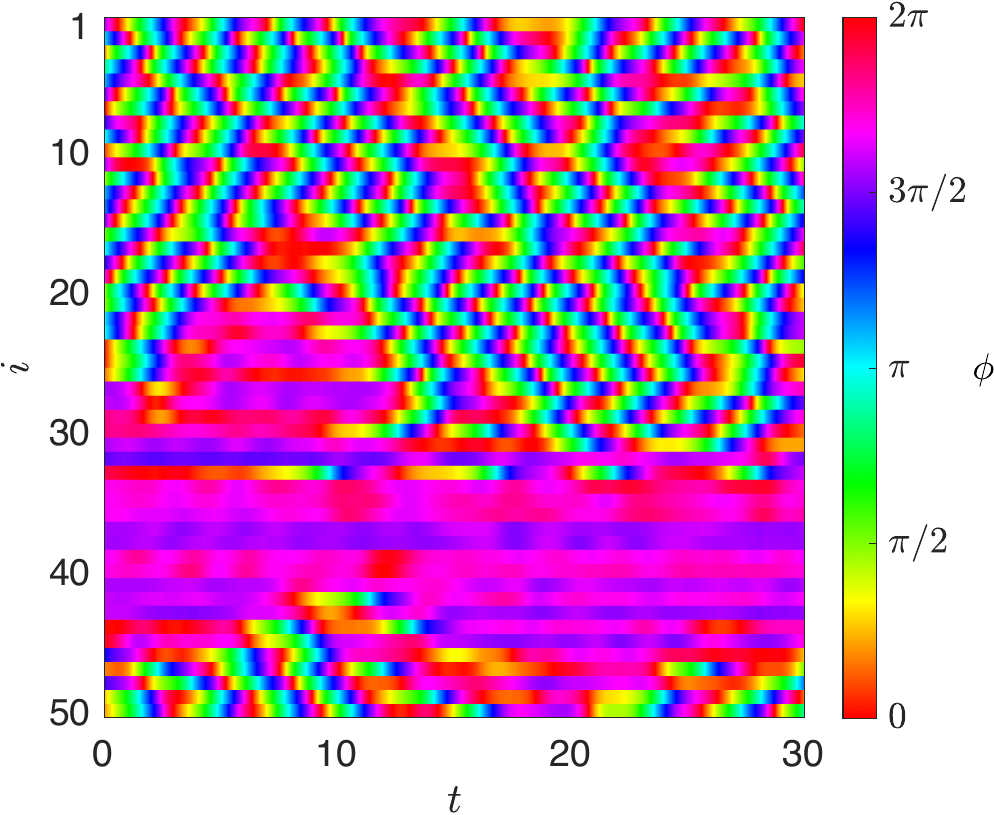}
\caption{
Time evolution of the $N=50$ phases $\phi_i$ of the theta neuron model (\ref{eq:theta_neurons}) showing a bump state. The theta neuron model parameters are $\kappa = 2$ and $\zeta_i \sim \mathcal{N}(-0.4,0.1)$, with connectivity matrix $B$ described in Section~\ref{sec:DA_theta}. 
}
\label{fig:theta_DA_1}
\end{figure}

The theta neuron model is a canonical toy model for neural activity in the brain. The model uses a SNIC-bifurcation normal form to model integrate-and-fire neurons \cite{Laing2014, Laing2018, LaingOmelchenko2020, OmelchenkoLaing2022}. 
In a network of $N$ theta neurons, the dynamics of the $i$-th neuron is given by
\begin{equation} \label{eq:theta_neurons}
\dot{\phi}_i = 1 - \cos \phi_i + \left( 1 + \cos \phi_i\right) \left(\zeta_i + \kappa I_i \right),
\end{equation}
where $\kappa$ is a coupling strength, $\zeta_i$ is an intrinsic parameter that represents the neuron's threshold energy to fire, and $I_i$ is the input from other neurons to neuron $i$ given by
\begin{equation} \nonumber
I_i = \frac{2\pi}{N} \sum_{j=1}^N B_{ij} P(\phi_j),
\end{equation}
where $P(\phi) = a (1 - \cos \phi)^2$, with $a$ such that $\int_0^{2\pi} P(\phi) d\phi = 2\pi$, and $B_{ij} \in \mathbb{R}$ is the connectivity between nodes $i$ and $j$, which may be negative for inhibitive coupling. When a neuron receives sufficient input from its neighbors, it overcomes its threshold energy and fires, completing a revolution of the unit circle. This firing provides input to the neuron's neighbors which in turn may cause them to fire. In such a network, many irregular and sustained firing patterns are possible depending on the model parameters and the initial condition. An example of a ``bump state'' is shown in Fig.~\ref{fig:theta_DA_1}, with a region of neurons that fire approximately periodically and another region of neurons that do not fire at all. Such bump states occur when there is short-range positive coupling, and long-range inhibitive coupling \cite{Bressloff2012}. Bump states are thought to be connected to short term memory \cite{WimmerEtAl2014}.


We will consider in Section~\ref{sec:DA_theta} the challenge to accurately estimate all the phases $\phi_i$ and all the intrinsic firing parameters $\zeta_i$, given only noisy observations of a subset of the phases and none of the firing parameters $\zeta_i$.

The true dynamics of the brain can be described by very high dimensional systems of delay differential equations, with a node for each of the billions of neurons. The dimensionality and complexity can be reduced by grouping sets of neurons into distinct regions of the brain, as well as simplifying the dynamics by ignoring, for instance, the delay. The theta neuron model (\ref{eq:theta_neurons}) is particularly simple and not very realistic, but we remark that slightly more complicated models, such as the FitzHugh-Nagumo model, are able to accurately replicate brain dynamic phenomena such as the temporary synchronization that occurs during epileptic seizures \cite{GersterEtAl20}. For brain dynamics, electroencephalography (EEG) can yield real-time data about the activity within different regions of the brain, and from that, the local phase and local frequency of the regions can be determined \cite{GersterEtAl20, WongEtAl23}.


\section{Ensemble Kalman Filter (EnKF) for coupled oscillators on networks} \label{sec:EnKF}

To solve the inference problem of estimating the state of a system and unknown parameters of the underlying model in the case that only some of the state variables are observed we employ the Ensemble Kalman Filter (EnKF). Kalman filters estimate the posterior distribution of the state given a model forecast and noisy observations. At each discrete time-step $t_n$, the EnKF combines a forecast ensemble $X_n^{\rm{f}}$ obtained by independently simulating the full non-linear model from a previous DA step, and noisy observations $Y_n$ to yield an improved `analysis' ensemble.

We describe here in detail the implementation of the EnKF for networks of coupled oscillators. We note that while many aspects of the method can be traced back to previous studies, changes must be made to accommodate for periodic phase variables that lie on a circle. We recall that the states are labelled as $\bphi\in \R^N$ and the parameters are $\bzeta \in \R^N$ (here $\bzeta$ represents either the natural frequencies $\omega_i$ of the Kuramoto model (\ref{eq:Kuramoto}) or the intrinsic parameters $\zeta_i$ of the theta neuron model (\ref{eq:theta_neurons})). To incorporate parameter estimation into an EnKF analysis step, an augmented state space  $\x  = (\bphi,\bzeta) \in \R^{2N}$ is used \cite{Anderson2001, Evensen2009a, Evensen2009}. 

In a Kalman filter, a Gaussian approximation of the posterior distribution is performed and one seeks expressions for the mean $\x^{\rm a}$ and the covariance matrix $P^{\rm a}$ of the posterior. We present here the algorithmic details to compute $\x^{\rm a}$ and $P^{\rm a}$. For more details and derivations the reader is referred to \cite{Evensen1994, Houtekamer1998, Jazwinski2007, Evensen2009a, Evensen2009, Reich2015, LawEtAl2015}. The analysis $\x^{\rm a}_n$ at time $t_n$ is an estimate of the state that combines a forecast $\x^{\rm f}_n$ and an observation $\yobs_n$. Treating $\x^{\rm f}_n$ and $\x^{\rm a}_n$, $n\ge 0$ as random variables and assuming a Gaussian distribution for $\x_{n+1}^{\rm f}$, the analysis step for the mean $\overline \x^{\rm a}_{n}$ is given by
\begin{align}
\overline \x^{\rm a}_{n} = \overline \x^{\rm f}_{n} - \K_{n} (H \overline \x^{\rm f}_{n}-\yobs_{n}),
\label{eq:KF1}
\end{align}
where the observation matrix $H =(H_\phi \; {\bf 0}) \in \mathbb{R}^{N_{\rm{obs}}\times 2N}$ projects from the (augmented) state space to the observation space. Here $N_{\rm{obs}}$ denotes the number of observed state variables. If, for example, only the first phase $\phi_1$ is observed, i.e $N_{\rm{obs}}=1$, then $H = (1,0,\cdots,0) \in \R^{1\times 2 N}$. The Kalman gain matrix $\K_n$ is given by
\begin{align} 
\label{eq:Kalman_gain}
\K_{n} = \Pf_{n} H^{T}\left(H \Pf_{n}H^{T}+{R}\right)^{-1}
\end{align}
with forecast covariance matrix $\Pf_{n}$ and observational error covariance matrix ${{R}}$. We assume a diagonal observational error covariance with ${R}=\eta^2 { I}$. The covariance matrix of the analysis is given by
\begin{equation} \label{eq:analysis_covariance}
P^{\rm a} = (I - KH)P^{\rm f}.
\end{equation}

We employ a stochastic EnKF \cite{BurgersEtAl98,Evensen2009a,Reich2015,LawStuart} to implement the Kalman analysis step. Ensemble Kalman filters allow for a dynamically adapted estimation of the forecast covariances $P^{\rm f}_n$, and they proved to be advantageous for nonlinear forward models, and for non-Gaussian augmented state variables. Consider an ensemble of states $\X \in \mathbb{R}^{2N\times M}$ consisting of $M$ members $\x^{(i)}\in \mathbb{R}^{2N \times 1}$, $i=1,\ldots,M$, that is,
\begin{equation}
\X=\left[ \x^{(1)},\x^{(2)},\dots,\x^{(M)} \right],
\end{equation}
with empirical mean
\begin{equation}
\overline{\x} = \frac{1}{M} \sum_{k=1}^M  \x^{(k)}.
\end{equation}
The associated matrix of ensemble deviations is given by
\begin{equation} \label{eq:ensemble_deviations}
\hat \X=\left[ \x^{(1)}-\overline{\x},\x^{(2)}-\overline{\x},\dots,\x^{(M)}-\overline{\x} \right].
\end{equation}
Note that for a $2\pi$-periodic phase variable $X_j=\phi_j$, the phase-mean $\bar{\phi}_j$ of the ensemble $\phi^{(k)}_j$, $k=1,\dots,M$, given by
\begin{equation} \label{eq:phase_mean}
\rho e^{i\bar{\phi}_j} = \frac{1}{M} \sum_{k=1}^M e^{i \phi^{(k)}_j},
\end{equation}
is used instead of the empirical mean and the ensemble deviations are defined as 
\begin{equation} \label{eq:phase_ensemble_deviations}
F\left(\phi_j^{(k)} - \bar{\phi}_j\right), \quad k=1,\dots,M,
\end{equation}
where $F(\theta) = \text{mod}(\theta + \pi, 2\pi) - \pi$ ensures that differences are centered at $0$ in the interval $[-\pi,\pi)$.
Ensembles for the forecast are denoted again by superscript f and those for the analysis by superscript a. In the forecast step the $\phi$-component of each ensemble member is propagated independently using the non-linear forecast model (i.e., either the Kuramoto model (\ref{eq:Kuramoto}) or the theta-neuron model (\ref{eq:theta_neurons})), updating the previous analysis ensemble $\X_{n-1}^{\rm a}$ to the next forecast ensemble $\X_{n}^{\rm f}$. The parameter component $\bzeta$ of each ensemble member is considered constant during the forecast step, i.e. $\dot\bzeta = 0$. The forecast covariance matrix $\Pf_{n}$ used in the analysis step (\ref{eq:KF1}) is estimated as a Monte-Carlo approximation from the forecast ensemble deviation matrix $\hat \X_{n}^{\rm f}$ via
\begin{align} \label{eq:P_empirical}
\Pf_{n} = 
\frac{1}{M-1}\hat \X_{n}^{\rm f}\,(\hat\X^{\rm f}_{n})^{\rm T}
\in \mathbb{R}^{2N\times 2N} .
\end{align}


To ensure that the analysis ensemble is statistically consistent with the Kalman filter, and, in particular, with the analysis error covariance (\ref{eq:analysis_covariance}),
in the stochastic ensemble Kalman filter \cite{BurgersEtAl98,Evensen2009a} observations $\yobs_{n}$ receive a stochastic perturbation $\boldsymbol{\eta}_{n}^{(i)}\in \R^{N_{\rm{obs}} \times 1}$, $i=1,\ldots,M$, drawn independently from the Gaussian observational noise distribution $\mathcal{N}({\bf 0},{R})$. The associated ensemble of perturbed observations $\tilde{\Z}_{n} \in \mathbb{R}^{N_{\rm{obs}} \times M}$ is given by
\begin{align}
\tilde{\Z}_{n} &= \left[ \yobs_{n} -  \boldsymbol{\eta}_{n}^{(1)},\yobs_{n} -\boldsymbol{\eta}_{n}^{(2)},
\ldots,
\yobs_{n} -  \boldsymbol{\eta}_{n}^{(M)} \right].
\end{align}
The EnKF analysis update step is then given by
\begin{align} 
\label{eq:EnKF}
\X_{n}^{\rm a} = \X_{n}^{\rm f} - K_{n} \Delta {\boldsymbol{I}}_{n},
\end{align}
with the Kalman gain defined by (\ref{eq:Kalman_gain}) using (\ref{eq:P_empirical}) and the stochastic innovation
\begin{align} 
\Delta {\boldsymbol{I}}_{n}=  H\X^{\rm f}_{n}-\tilde{\Z}_{n}.
\end{align}
Finite ensembles are prone to an underestimation of error covariances which may cause filter divergence in which the filter believes its own forecast since $ \Pf $ is small, and the forecast is not corrected by new incoming observations. To mitigate against such finite ensemble size effects we employ, as is typically done, covariance inflation with $\Pf_n \to \delta \Pf_n$ with $\delta=1.001$ \cite{Anderson1999}. This multiplicative inflation does not affect the ensemble mean but increases the forecast error covariance. The exact causes of how finite ensemble sizes lead to underestimation of covariances is complex \cite{Houtekamer1998, vanLeeuwen99, SakovBartello08, Snyder14}. In chaotic dynamical systems the underestimation of error covariances may be further exaggerated by ensemble members aligning, on short time scales, with the most unstable direction \cite{NgEtAl11}. 


\subsection{Splitting the analysis step into phase and parameter components} \label{sec:DA_splitting}

For our ensemble $\X=\left[ \x^{(1)},\x^{(2)},\dots,\x^{(M)} \right]\in \mathbb{R}^{2N\times M}$, we can write each ensemble member in the form
\begin{equation}
X^{(j)} = \begin{pmatrix}
\bphi^{(j)} \\
\bzeta^{(j)}
\end{pmatrix} \in \mathbb{R}^{2N\times 1},
\end{equation}
such that $\bphi^{(j)} = \left(\phi_1^{(j)},\dots,\phi_N^{(j)}\right)^T \in \mathbb{R}^{N\times 1}$ and $\bzeta^{(j)} = \left(p_1^{(j)},\dots,p_N^{(j)}\right)^T\in \mathbb{R}^{N\times 1}$. Following (\ref{eq:ensemble_deviations})-(\ref{eq:phase_ensemble_deviations})
we can define ensemble deviations for the phases and parameters separately, i.e.,
\begin{equation}
\hat{\bm{\Phi}} = \left[ F\left(\bphi^{(1)} - \bar\bphi\right), \dots, F\left(\bphi^{(M)} - \bar\bphi\right) \right], \quad \hat{\bm{\Pi}} = \left[ \bzeta^{(1)} - \bar\bzeta, \dots, \bzeta^{(M)} - \bar\bzeta \right],
\end{equation}
where $\bar\bphi$ is the vector of phase means (\ref{eq:phase_mean}) and $F$ is as in (\ref{eq:phase_ensemble_deviations}). Therefore, the ensemble deviations (\ref{eq:ensemble_deviations}) are given by
\begin{equation}
\hat \X = \begin{pmatrix}
\hat{\bm\Phi} \\
\hat{\bm\Pi}
\end{pmatrix}.
\end{equation}
From (\ref{eq:P_empirical}), it follows that the forecast covariance matrix is given by
\begin{equation}
\Pf_n = \frac{1}{M-1}\begin{pmatrix}
\hat{\bm\Phi}_n^{\rm{f}} (\hat{\bm\Phi}_n^{\rm{f}})^T & \hat{\bm\Phi}_n^{\rm{f}} (\hat{\bm\Pi}_n^{\rm{f}})^T \\
\hat{\bm\Pi}_n^{\rm{f}} (\hat{\bm\Phi}_n^{\rm{f}})^T & \hat{\bm\Pi}_n^{\rm{f}} (\hat{\bm\Pi}_n^{\rm{f}})^T
\end{pmatrix} = 
\begin{pmatrix}
\Pf_{\bphi \phi} & \Pf_{\bphi \bzeta} \\
\Pf_{\bzeta \bphi} & \Pf_{\bzeta \bzeta}
\end{pmatrix},
\end{equation}
such that $\Pf_{\bphi \bphi}$ is the covariance matrix of the phases only, $\Pf_{\bphi \bzeta}$ is the covariance between the phases and parameters, etc. 

Since we only observe the phases $\phi$, we can separate the state and parameter update of the Kalman analysis step (\ref{eq:KF1}) as
\begin{subequations}
\begin{align}
\overline \bphi^{\rm a}_{n} &= \overline \bphi^{\rm f}_{n} -  \Pf_{\bphi\bphi} H_{\bphi}^T \left( H_{\bphi} \Pf_{\bphi\bphi} H_{\bphi}^T+ R \right)^{-1}\Delta \boldsymbol{I}_{n}\\
\overline \bzeta^{\rm a}_{n} &= \overline \bzeta^{\rm f}_{n} - \Pf_{\bzeta\bphi} H_{\bphi}^T \left( H_{\bphi} \Pf_{\bphi\bphi} H_{\bphi}^T+{ R}\right)^{-1}\Delta \boldsymbol{I}_{n} \label{eq:parameter_update}
\end{align}
\label{eq:KF2}
\end{subequations}
with innovation
\begin{align}
    \Delta \boldsymbol{I}_{n} := H_{\bphi} \overline \bphi_{n}^{\rm f}-\yobs_{n}.
\end{align}
We note that the covariance matrix $\Pf_{\bzeta \bzeta}$, between the parameters and themselves, does not enter this equation. 


\subsection{Network specific localization} \label{sec:spurious_loc}

A source of error in ensemble-based methods arises due to spurious correlations caused by the finite size of the ensemble. If two variables are not correlated, the Monte-Carlo approximation of the covariance yields entries of ${\mathcal{O}(1/\sqrt{M}})$ compared to the true entry which is $0$. One could reduce the effect of spurious correlations by using a very large ensemble size $M$, but that is computationally prohibitive for high-dimensional systems. An alternative approach is so-called	``localization'', whereby topological information is used to predetermine and suppress spurious correlations \cite{Houtekamer2001}, which allows for a smaller ensemble to be used. For example, in weather forecasting, correlations between geographically distant grid-points should be very small and any large correlations are deemed spurious. It is less clear how to best incorporate localization for dynamics on networks. 

We use the Kuramoto model (\ref{eq:Kuramoto}) as an illustrative example of spurious correlations, which guides the appropriate form of localization that should be performed. For illustrative purposes we assume that all phase variables $\phi_j$ are observed at each time step, and that the oscillators are arranged in a ring network topology with coupling radius $r=3$, where each node is connected to its $2r$ nearest neighbors. 

\begin{figure*}[tbp]
\centering
\includegraphics[width=\textwidth]{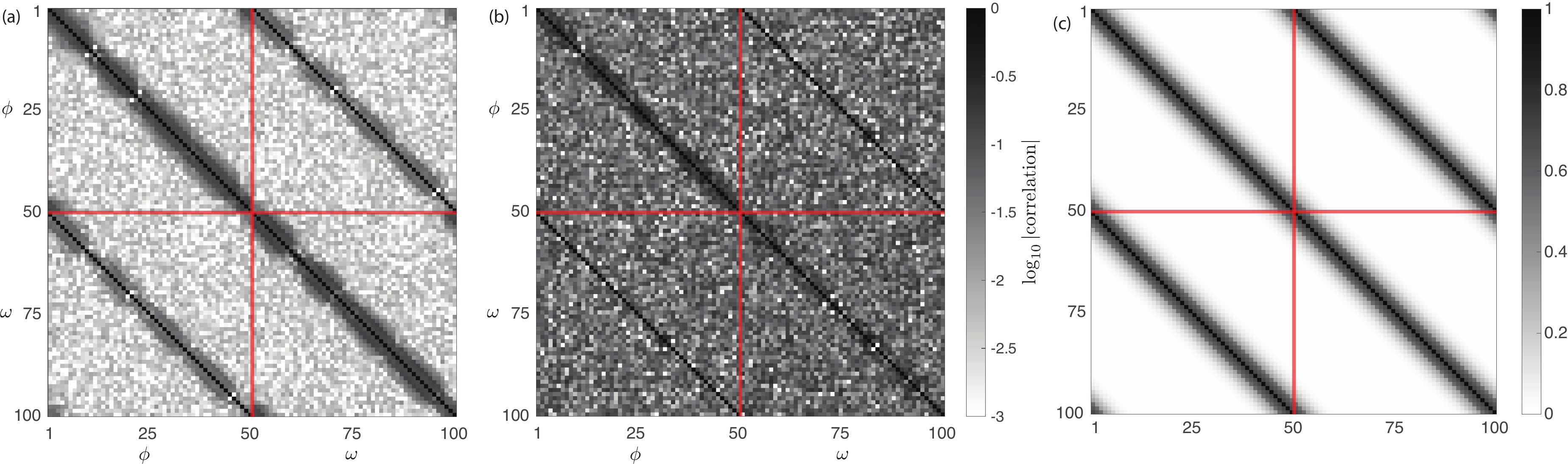}
\caption{
(a)~Time-averaged correlation matrix $Q$ for a large ensemble ($M = 10100$), and (b)~for a small ensemble ($M=101$), obtained by applying the standard EnKF to the Kuramoto model (\ref{eq:Kuramoto}). The time-averaging is performed for a single realization. The augmented state space is $(\phi_{1,\dots,50},\omega_{1,\dots,50})$, hence the block matrix substructure (indicated by red lines) in $Q$ with $\phi$-$\phi$ correlations, $\phi$-$\omega$ correlations, $\omega$-$\phi$ correlations and $\omega$-$\omega$ correlations. Note that the absolute values of the correlations are shown on a logarithmic scale to more clearly illustrate the differences in magnitude between the smaller correlations. The Kuramoto model parameters are as in Fig.~\ref{fig:KM_phi_of_t}. (c)~The localization matrix $\mathcal{L}$ (\ref{eq:localization_matrix}) for the associated ring topology using $\lambda = 0.46$ as determined in Section~\ref{sec:choosing_lambda}.
}
\label{fig:KM_DA_corr}
\end{figure*}

To demonstrate spurious correlations, we normalize the covariance matrices $P^{\rm{f}}_n$ at each DA step to obtain correlation matrices $Q_n$. We generate noisy observations $Y_n$ by adding Gaussian noise with covariance $R = \eta^2 I$ (here $\eta = 0.02$) to simulated trajectories of the Kuramoto model (\ref{eq:Kuramoto}). Fig.~\ref{fig:KM_DA_corr}(a,b) shows the averaged correlation matrix $Q$,  for a large ensemble $M = 100(2N + 1)=10,100$ and a small ensemble $M=2N+1=101$\footnote{Since the augmented state space is $2N$-dimensional, the minimum ensemble size to ensure a full rank covariance matrix is $M=2N + 1$.}. For the large ensemble, correlations decay rapidly away from the leading diagonal in each sub-block (Fig.~\ref{fig:KM_DA_corr}(a)), reflecting the underlying ring topology. For the small ensemble, however, there are large spurious correlations far from the diagonal (Fig.~\ref{fig:KM_DA_corr}(b)), wrongly suggesting connections between poorly connected nodes. Moreover, the information about the network topology is completely lost in the $\phi$-$\omega$ correlations $P^{\rm f}_{\bphi \bzeta}$, which are crucial for estimating the parameters $\omega_i$ (cf. (\ref{eq:parameter_update})). To suppress these spurious correlations, and allow for smaller ensembles to be used, localization is required, where at each DA step the covariance matrix $P^{\rm{f}}_n$ is replaced by
\begin{equation} \nonumber
\tilde{P}^{\rm{f}}_n = \mathcal{L} \circ P^{\rm{f}}_n,
\end{equation}
where $\circ$ denotes the Schur (elementwise) product, and $\mathcal{L}$ is the localization matrix. 
Standard localization functions, such as Gaspari-Cohn localization \cite{Houtekamer2001,GaspariCohn1999}, are not suitable for networks, because they cannot be embedded in $\mathbb{R}^n$ for any $n$. We discuss this in detail in Section~\ref{sec:Gaspari-Cohn}. 
Instead we require $\mathcal{L}$ to have the following properties:
\begin{itemize}
	\item $\mathcal{L}$ encodes the network connectivity, i.e., $\mathcal{L}_{ij}$ is large if nodes $i$ and $j$ are well connected in the sense that they are direct neighbors and/or there are many paths of short length connecting them, and $\mathcal{L}_{ij}$ is small if nodes $i$ and $j$ are poorly connected,
	\item $\mathcal{L}$ is a correlation matrix, i.e., symmetric, positive semi-definite and $\mathcal{L}_{ii}=1$.
\end{itemize}
As a natural choice of a matrix that has the two above properties, we propose the matrix exponential of the network adjacency matrix.
Consider the (symmetric positive semi-definite matrix) covariance matrix, $\mathcal{E} = \exp(\lambda A)$, where $\lambda$ is a tunable parameter and $A$ is the network adjacency matrix (assumed to be symmetric). The matrix
\begin{equation} \label{eq:mat_exp}
\mathcal{E} = \exp(\lambda A) = \sum_{n=0}^\infty \frac{\lambda^n}{n!} A^n
\end{equation}
is a weighted sum, with decreasing weights, of powers $A^n$. The entries $(A^n)_{ij}$ count the number of paths of length $n$ between nodes $i$ and $j$. Hence, $\mathcal{E}$ encodes the network topology. 
The entry $\mathcal{E}_{ij}$ quantifies the connectedness between nodes $i$ and $j$; $\mathcal{E}_{ij}$ is large if nodes $i$ and $j$ are directly connected, or connected via many paths with short length, and $\mathcal{E}_{ij}$ is close to zero otherwise.
A correlation matrix $L$ is obtained by normalizing $\mathcal{E}$ according to
\begin{equation} \label{eq:localization_matrix_0}
L = \left( \text{diag}(\mathcal{E}) \right)^{-\frac{1}{2}} \mathcal{E} \left( \text{diag}(\mathcal{E}) \right)^{-\frac{1}{2}}.
\end{equation}
To account for the augmented state space with both phases and frequencies we introduce the block matrix
\begin{equation} \label{eq:localization_matrix}
\mathcal{L} =  \begin{pmatrix} 
L & L \\ L & L
\end{pmatrix}.
\end{equation}
An example of $\mathcal{L}$ is shown in Fig.~\ref{fig:KM_DA_corr}(c) for the ring topology with $N=50$, $r=3$ and $\lambda = 0.46$. The parameter $\lambda$ controls the rate of decay for distant nodes and is chosen to suppress spurious correlations. We discuss the appropriate choice of $\lambda$ and a simple heuristic method for determining $\lambda$ in Section~\ref{sec:choosing_lambda}.

We will show in Section~\ref{sec:results} that introducing this network-specific localization $\mathcal{L}$ significantly improves the prediction and estimation accuracy of the EnKF with partial observations for both the Kuramoto model (\ref{eq:Kuramoto}) and networks of theta neurons (\ref{eq:theta_neurons}).

We remark that there exist other localization matrices that satisfy the two desired properties above, and we discuss some alternative localization matrices in Section~\ref{sec:Gaspari-Cohn}.

\subsection{Choosing the localization parameter $\lambda$} \label{sec:choosing_lambda}

The parameter $\lambda$ in the localization matrix $L$ (\ref{eq:mat_exp}), (\ref{eq:localization_matrix_0}) is important for controlling the rate of decay of correlations for distant nodes. The optimal value of $\lambda$ could be found through numerical optimization, minimizing the RMS error of the DA algorithm across many simulations. Here we present a cheap heuristic method to choose a value of $\lambda$ at first for ring network topologies, and then we will generalize the method to random network topologies. 


We note that the distance at which correlations become spurious, and, hence, the optimal choice of $\lambda$, will depend on the size of the ensemble $M$, such that larger ensembles are able to better estimate the weaker long range correlations \cite{Houtekamer1998}. The heuristic method that we present here is based on the typical decay of correlations as observed from DA for a large ensemble, which allows for a reliable estimation of the typical distance at which correlations become weak.

\subsubsection{Ring network topologies} \label{sec:choosing_lambda_ring}

\begin{figure*}[tbp]
\centering
\includegraphics[width=0.5\textwidth]{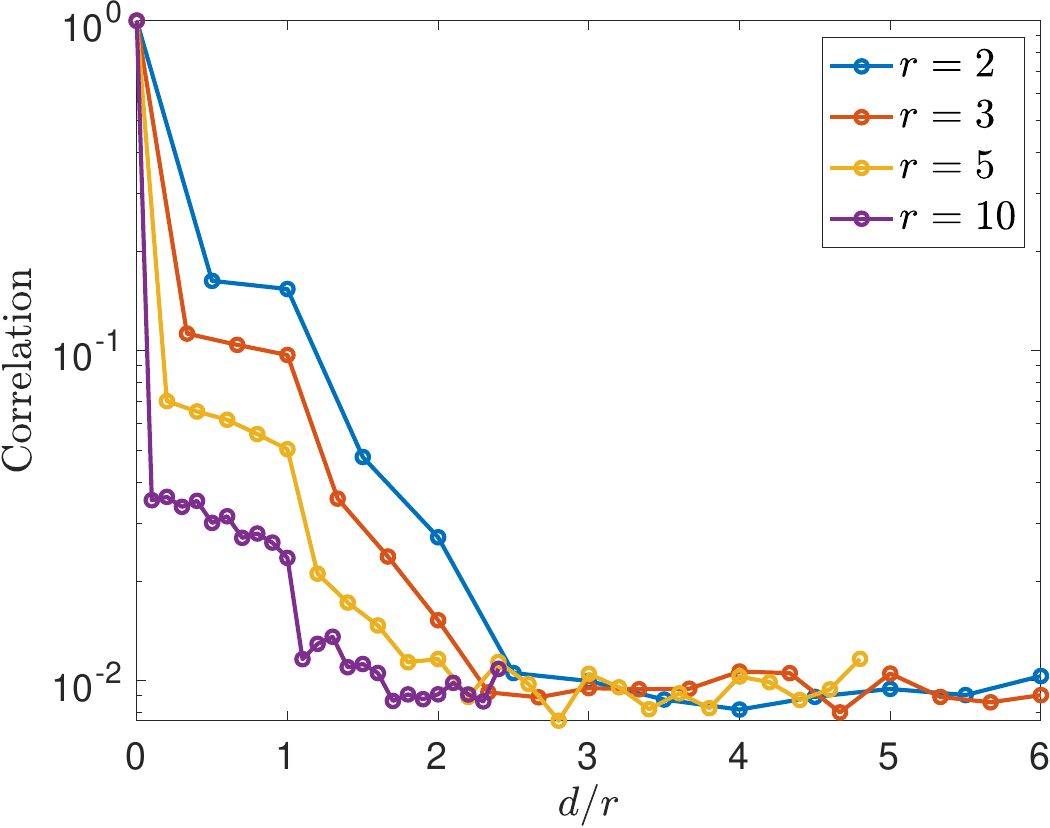}
\caption{
Average forecast correlation (averaging $P_k^{\rm f}$ over all nodes and all time steps) as a function of distance $d$ from the node, scaled by the coupling radius $r$, for $r=2,3,5,10$. All results are obtained from runs of the standard EnKF applied to the Kuramoto model ($N=50$, $\kappa=80/r$, $\omega_i\sim \mathcal{N}(0,0.1)$ with a ring topology), with a large ensemble $M=100(2N+1)$. Note that a coupling strength $\kappa = 80/r$ is used to maintain a constant effective coupling strength $\kappa/\langle k \rangle$, where $\langle k \rangle$ is the mean degree, which ensures that the dynamics of the Kuramoto model are equivalent for all $r$.
}
\label{fig:KM_DA_corr_vs_i}
\end{figure*}

\begin{figure*}[tbp]
\centering
\includegraphics[width=0.4\textwidth]{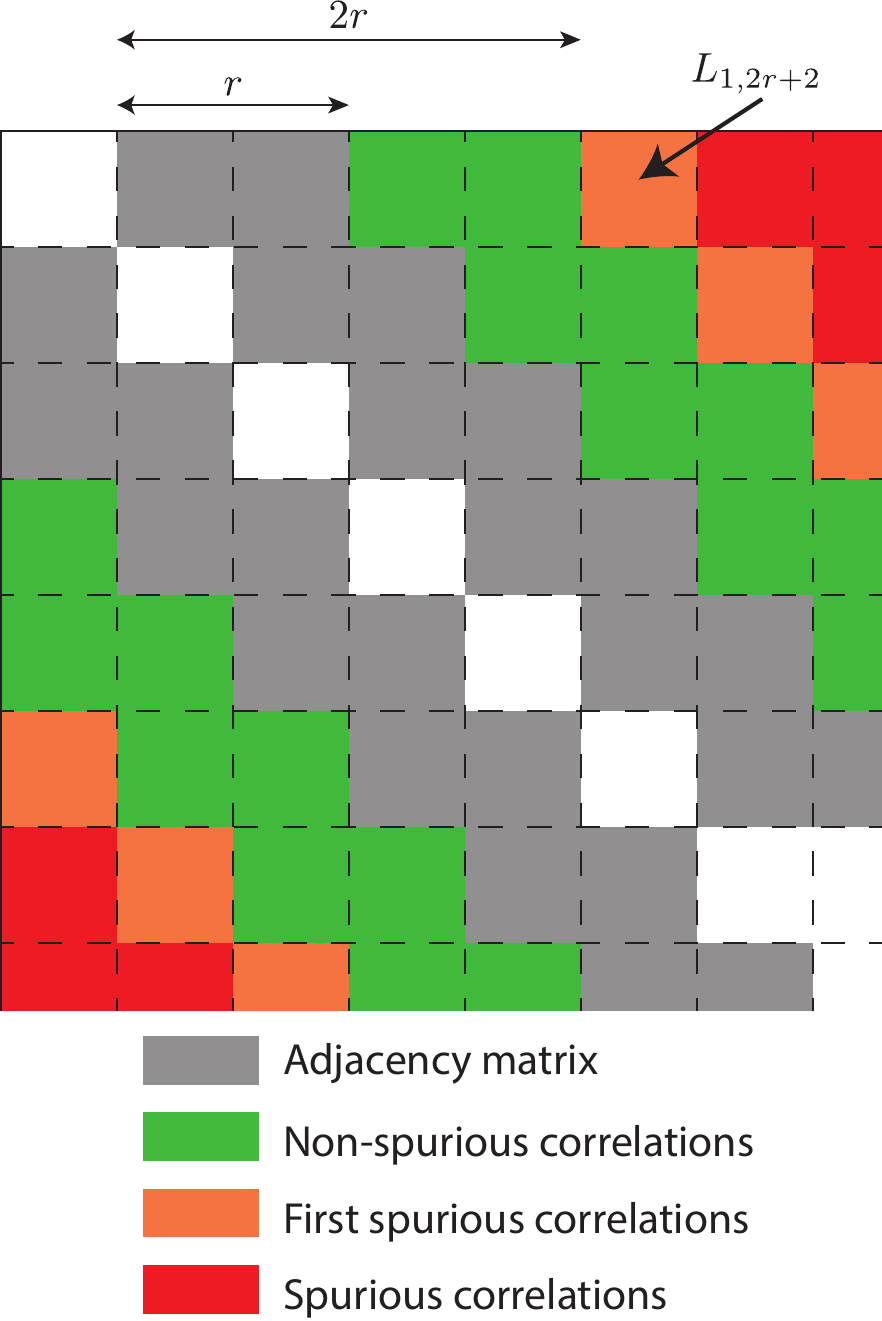}
\caption{
Schematic diagram for the condition (\ref{eq:ring_lambda_1}) that defines $\lambda$. Shown is the top right corner of the matrix $L$, the gray squares indicate the underlying adjacency matrix (here a ring with $r=2$), the green squares indicate correlations that should be considered significant because they are within a distance of $2r$ from the node, and orange/red squares indicate correlations that should be considered weak and potentially spurious because the distance from the node is greater than $2r$. The orange squares are those whose value will be $\epsilon$ for $\lambda$ satisfying (\ref{eq:ring_lambda_1}) with equality, e.g., in the first row $L_{1,2r+2}=\epsilon$.
}
\label{fig:lambda_schematic}
\end{figure*}

 Considering a ring topology with connectivity radius $r$, we find that the correlations for large ensembles are weak beyond a radius of $2r$. That is, for a node $j$, only the nodes $k$ that have shortest path length from $j$ to $k$ less than or equal to $2$ have significant correlations. This is shown in Fig.~\ref{fig:KM_DA_corr_vs_i}, which shows the average correlation as a function of distance from the node, scaled by the connectivity radius $r$. This data is obtained from the standard EnKF applied to the Kuramoto model with $N=50$ and with a large ensemble $M=100(2N+1)$, as in Fig.~\ref{fig:KM_DA_corr}(a) which shows correlations for $r=3$. Fig.~\ref{fig:KM_DA_corr_vs_i} shows that correlations become weak beyond a distance of $2r$ ($d/r=2$).
The rapid decay of correlation that we have observed is expected due to the pairwise coupling of nodes in both the Kuramoto model (\ref{eq:Kuramoto}) and theta neuron model (\ref{eq:theta_neurons}), which restricts the flow of information.
  Using this information, considering node $j=1$, the nodes within a radius of $2r$, and which will have significant correlations are $k=1,\dots,2r+1$ and $k=N-r + 1,\dots,N$, and all other nodes, i.e., $k = 2r+2,\dots,N-r$ should have weak correlations. However, these nodes that should have weak correlations may in fact have large, spurious, correlations due to a small ensemble size $M$. These potentially spurious correlations are those which should be suppressed. Knowing which correlations we wish to suppress, the next question is how much do we need to suppress them? For this, we choose an $0<\epsilon < 1$, which will be the upper bound for the entries in $L$ (\ref{eq:localization_matrix_0}) beyond a radius of $2r$ from each node. Here we have chosen $\epsilon = 0.1$. Due to the symmetry of the ring topology, and the fact that $\exp(\lambda A)$ decays away from its diagonal, we require that
\begin{equation} \label{eq:ring_lambda_1}
L_{1,2r+2} = \frac{\mathcal{E}_{1,2r+2}}{\mathcal{E}_{1,1}}  \leq \epsilon,
\end{equation}
where $\mathcal{E}=\exp(\lambda A)$ as in (\ref{eq:mat_exp}). Changing the inequality in (\ref{eq:ring_lambda_1}) to an equality gives an equation that can be solved (numerically) for $\lambda$ in terms of $r$ and $\epsilon$. This is demonstrated schematically in Fig.~\ref{fig:lambda_schematic}, where the orange squares are the first weak correlations, with $L_{ij}=\epsilon$. In particular, in the first row we have $L_{1,2r+2}=\epsilon$.

\begin{figure*}[tbp]
\centering
\includegraphics[width=\textwidth]{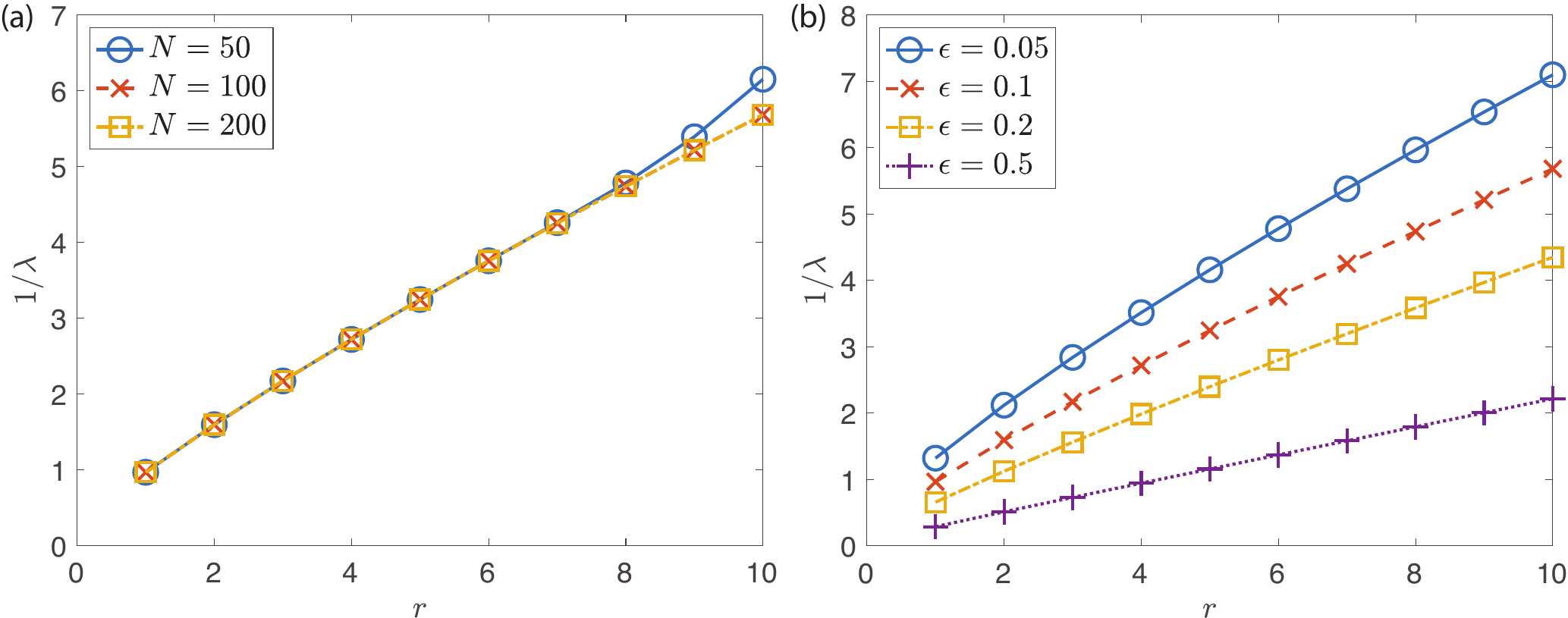}
\caption{
$1/\lambda$ as a function of $r$ for various values of the number of oscillators $N$ and threshold $\epsilon$. In all cases $\lambda$ is obtained from (\ref{eq:ring_lambda_1}). (a)~Fixed $\epsilon=0.1$, $N=50, 100, 200$. (b)~Fixed $N=200$, $\epsilon=0.05,0.1,0.2,0.5$.
}
\label{fig:lambda_vs_i}
\end{figure*}

For a fixed value of $\epsilon$, we find that $1/\lambda$ scales approximately linearly with $r$, independent of $N$ provided $r \ll N$. This is shown in Fig.~\ref{fig:lambda_vs_i}. Fig.~\ref{fig:lambda_vs_i}(a) shows that for fixed $\epsilon=0.1$, the value of $\lambda$ obtained from (\ref{eq:ring_lambda_1}) does not depend on $N$, provided that $r \ll N$. Fig.~\ref{fig:lambda_vs_i}(b) shows different approximately linear scalings between $1/\lambda$ and $r$ for various values of $\epsilon$ (all using $N=200$). This approximately linear scaling will now be utilized to determine appropriate values of $\lambda$ for general network topologies.

\subsubsection{Complex network topologies}  \label{sec:choosing_lambda_random}

We determine the value of $\lambda$ to use in the localization matrix (\ref{eq:mat_exp}), (\ref{eq:localization_matrix_0}) for a complex network topology based on a ring topology with equivalent mean degree. Consider a complex network topology with mean degree $\langle k \rangle$. The ring topology with the same mean degree has connectivity radius $r^* = \langle k \rangle/2$ (ignoring for now that this is likely not an integer). We first find $\lambda^*$ corresponding to the ring topology with $r = r^*$.  If $r^*$ is an integer, we follow the procedure described in the previous section for the ring topology, otherwise, we interpolate to find $\lambda^*$. We could linearly interpolate between the two nearest integer values of $r$, but we can do better by noting that $1/\lambda$ scales approximately linearly with $r$, as shown in Fig.~\ref{fig:lambda_vs_i}. Thus, we find the fit of the form
\begin{equation}
\frac{1}{\lambda} = m r + c,
\end{equation}
passing through $(r_1,1/\lambda_1)$ and $(r_2,1/\lambda_2)$, where $r_1 = \lfloor r^* \rfloor \leq r^*$ and $r_2 = \lceil r^* \rceil\geq r^*$ are the two nearest integers to $r^*$, and $\lambda_1$ and $\lambda_2$ are the corresponding values of $\lambda$ obtained from ring topologies. That is,
\begin{equation} \label{eq:ER_lambda_1}
m = \frac{1/\lambda_2 - 1/\lambda_1}{r_2 - r_1}, \quad c=\frac{1}{\lambda_1} - m r_1.
\end{equation} 
The value $\lambda^*$ is then given by
\begin{equation} \label{eq:ER_lambda_2}
\lambda^* = (m r^* + c)^{-1}.
\end{equation}

As an example, for an Erd\H{o}s-R\'{e}nyi graph $N=50$ nodes and coupling probability $p=0.1$, such that each pair of nodes are coupled with probability $p$, the mean degree is $\langle k \rangle = N(p-1) = 4.5$. The equivalent ring topology has $r^* = 2.45$. The two nearest integers are $r_1 = 2$ and $r_2 = 3$. From the method for ring topologies in the previous section we obtain $\lambda_1 = 0.627$ and $\lambda_2 = 0.460$. Substituting these values into (\ref{eq:ER_lambda_1}), (\ref{eq:ER_lambda_2}) yields $\lambda^* = 0.539$, which is the value of $\lambda$ used in Section~\ref{sec:DA_KM}\ref{sec:DA_KM_ER} for random Erd\H{o}s-Renyi networks.

\subsection{Alternative localization matrices} \label{sec:Gaspari-Cohn}

Localization has been widely employed in applications of the EnKF. However, most applications assume an underlying Euclidean geometry such that the distance between nodes is their Euclidean distance. Based on the assumption of Euclidean geometry, methods have been created to generate localization functions. A widely used method is that of Gaspari \& Cohn \cite{GaspariCohn1999}, with the localization function
\begin{equation} \label{eq:GC_function}
C(z,c) = \begin{cases}
-\frac{1}{4} \left(\frac{|z|}{c}\right)^5 + \frac{1}{2} \left( \frac{z}{c} \right)^4 + \frac{5}{8} \left(\frac{|z|}{c}\right)^3 -  \frac{5}{3} \left( \frac{z}{c} \right)^2 + 1, & 0 \leq |z| \leq c, \\
\frac{1}{12} \left(\frac{|z|}{c}\right)^5 - \frac{1}{2} \left( \frac{z}{c} \right)^4 + \frac{5}{8} \left(\frac{|z|}{c}\right)^3 + \frac{5}{3} \left( \frac{z}{c} \right)^2 - 5 \left(\frac{|z|}{c}\right) + 4 - \frac{2}{3} \left(\frac{|z|}{c}\right)^{-1}, & c\leq |z| \leq 2c, \\
0, & 2c \leq |z|,
\end{cases}
\end{equation}
where $|z|$ is the Euclidean distance between the nodes and $c$ is a length-scale that controls the decay of correlation. The function $C$ is a $5$th-order piecewise rational function with compact support. In generating a localization matrix $L$, one would define $L_{ij} = C(|x_i - x_j|,c)$, where $x_i$ is the position of node $i$ in space, and $|x|$ denotes the Euclidean norm. Note, the proof that $C$ is a correlation function relies on the Euclidean distance being used.

\begin{figure*}[tbp]
\centering
\includegraphics[width=0.5\textwidth]{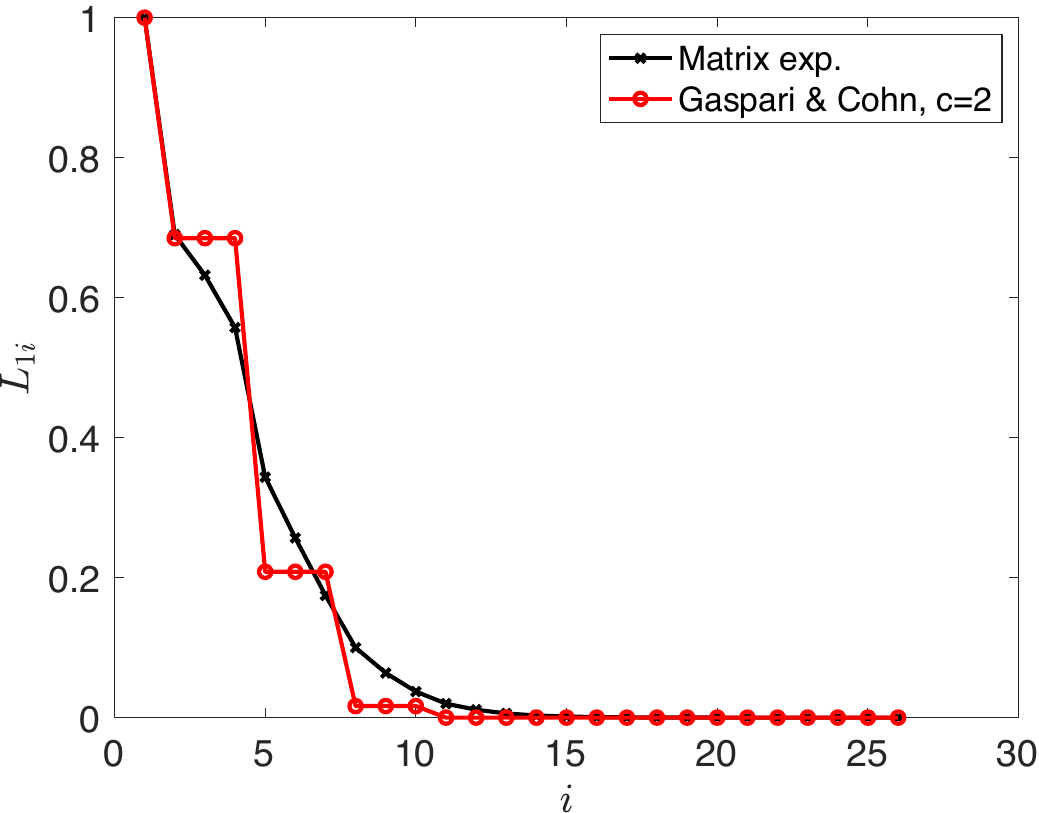}
\caption{
Values $L_{1i}$ from the first row of the matrix exponential correlation matrix shown in Fig.~\ref{fig:KM_DA_corr}(b) using (\ref{eq:localization_matrix_0}) with $\lambda = 0.46$  (black crosses) and values $\tilde{L}_{1i} = C(d(1,i),c)$ using the Gaspari-Cohn function with shortest path length distance (\ref{eq:GC_function}) with $c=2$ (red circles). Both use an underlying ring topology with $N=50$ and $r=3$.
}
\label{fig:matexp_vs_GC}
\end{figure*}

When considering localization on networks, there are many possible definitions of the ``distance'' between two nodes. Possibly the simplest definition is the shortest path length, that is, the distance $d(i,j)$ between nodes $i$ and $j$ is the length of the shortest path connecting $i$ to $j$. With this definition, one can, in principle, define the function $C$ from (\ref{eq:GC_function}), and create a matrix $\tilde{L}_{ij} = C(d(i,j),c)$, where here $c$ is a characteristic length for meaningful correlations. However, this construction does not in general yield a correlation matrix, in particular, the matrix $\tilde{L}$ is not always positive semi-definite, which may cause the EnKF to ``blow up'' due to the term $(H \Pf_{n}H^{T} + R)^{-1}$ in the Kalman gain (\ref{eq:Kalman_gain}). Nevertheless, we can compare the matrix $\tilde{L}$ with $L$ obtained from (\ref{eq:localization_matrix_0}). In Section~\ref{sec:choosing_lambda} we noted that for ring topologies correlations appear to be weak for node distances greater than $2$, so a choice $c\approx 2$ is natural. Fig.~\ref{fig:matexp_vs_GC} shows the first rows of the respective localization matrices $L$ and $\tilde{L}$ for the ring topology with $N=50$ and $r=3$ using the matrix exponential construction (\ref{eq:localization_matrix_0}) (black crosses) compared to the Gaspari-Cohn construction with shortest path length distance (red circles). There is very good agreement between the two constructions. 


The similarity between the matrix exponential and the Gaspari-Cohn function with shortest path length distance (\ref{eq:GC_function}) suggests that the matrix exponential is a good choice for localization. One advantage of the Gaspari-Cohn construction (\ref{eq:GC_function}) is that the function is compactly supported, resulting in a sparse localization matrix, which in turn reduces the computational cost of data assimilation. One could produce a sparse matrix from the localization matrix $L$ derived from the matrix exponential (\ref{eq:localization_matrix_0}) by setting all entries below a threshold value equal to zero, however, the resulting matrix is not guaranteed to be positive semi-definite. Alternatively, a sparse localization matrix can be obtained by considering a normalization of a matrix of the form
\begin{equation}
B = \sum_{i=0}^n \alpha_i A^i,
\end{equation}
which is positive definite provided that $A$ is symmetric and $\sum_{i=0}^n \alpha_n \eta^i\geq 0$ for all eigenvalues $\eta$ of $A$, and is sparse provided that $n \ll N$. This construction allows for $n+1$ tunable parameters $\alpha_i$. For example, using the matrix exponential as a guide, one could consider the truncation
\begin{equation}
B = \sum_{i=0}^n \frac{\lambda^i}{i!} A^i,
\end{equation}
for $n \ll N$ and with one parameter $\lambda$. However, one would have to check that $\sum_{i=0}^n \frac{\lambda^i}{i!} \eta^i$ is non-negative for all eigenvalues $\eta$ of $A$. Consideration of sparse localization matrices of this form is left as an avenue for future work.

\section{Results} \label{sec:results}

Each of our data assimilation trials are initialized with an initial ensemble $\X_0^{\rm a}$ at $t=0$ with $M=2N+1$ members and $\bphi_0^{\rm a}\sim {\mathcal N}(\bphi_0 + \bm{\eta}_\phi, \sigma_{\phi,1}^{2} I)$, such that $\bphi_0$ is the true value at $t=0$ and $\bm{\eta}_\phi \sim \mathcal{N}(0,\sigma_{\phi,2}^2) \in \mathbb{R}^N$. In essence, $\sigma_{\phi,2}$ controls the perturbations of the ensemble means away from the truth, and $\sigma_{\phi,1}$ controls the spread of the ensemble around the ensemble means. The initial ensemble for the parameters $\bzeta$ has the analogous form $\bzeta_0^{\rm a} \sim {\mathcal N}(\bzeta_0 + \bm{\eta}_p, \sigma_{p,1}^{2} I)$, such that $\bm{\eta}_p \sim \mathcal{N}(0,\sigma_{p,2}^2) \in \mathbb{R}^N$.  For data assimilation with the Kuramoto model we use $\sigma_{\phi,1}^2 = \sigma_{\phi,2}^2 = 0.25$, and $\sigma_{\omega,1}^2 = \sigma_{\omega,2}^2 = 0.25\, \sigma_\omega^2 =  0.025$, where $\sigma_\omega^2 = 0.1$ is the variance of the natural frequency distribution. For data assimilation with the theta neuron network we use $\sigma_{\phi,1}^2 = \sigma_{\phi,2}^2 = 0.04$, and $\sigma_{\zeta,1}^2 = \sigma_{\zeta,2}^2 = 0.04\, \sigma_\zeta^2 =  0.004$, where $\sigma_\zeta^2 = 0.1$ is the variance of the intrinsic firing parameter distribution. 
 The ensemble size $M=2N+1$ is chosen since it is the smallest ensemble size ensuring that the covariance matrices have full rank. This eliminates any problems stemming from rank-deficiency of the covariances (see \cite{Snyder14}) and allows us to focus exclusively on the effect of the topological information of the localization.

To test our data assimilation methods we generate noisy observations $Y_n$ by first simulating the model and then artificially adding noise with prescribed covariance ${{R}} = \eta^2I$ with $\eta = 0.02$. To quantify the skill of the data assimilation procedure, we compute the root mean square (RMS) errors between the true and the DA analysis mean values for the phase variables and the parameters. The RMS error for the phases $\phi_i$ at time $t$ is
\begin{equation} \label{eq:rms_phi}
E_{\phi}(t) = \frac{1}{\sqrt{N}} \sqrt{ \sum_{i=1}^N \left(\phi_i(t) - \phi^{\rm a}_i(t) \right)^2},
\end{equation}
where $\phi^{\rm a}_i(t)$ is the DA estimated value (the analysis mean) and $\phi_i(t)$ is the true value. As in (\ref{eq:phase_ensemble_deviations}), we use the function $F$ to center the phase differences around zero, i.e., in the range $[-\pi,\pi)$. The RMS error for the model parameters, either the natural frequencies $\omega_i$ of the Kuramoto model (\ref{eq:Kuramoto}) or the intrinsic firing parameters $\zeta_i$ of the theta neuron model (\ref{eq:theta_neurons}), are computed analogously, and denoted by $E_\omega(t)$ and $E_\zeta(t)$, respectively.

\subsection{Data assimilation for the Kuramoto model}  \label{sec:DA_KM}

We return to the question posed previously: Can DA accurately predict and estimate all of the phase variables $\phi_i$ and natural frequencies $\omega_i$ in the Kuramoto model (\ref{eq:Kuramoto}) if only a subset of the phases are observed, with none of the frequencies being observed?

We will first show that our localization scheme significantly improves the accuracy of the EnKF when applied to the Kuramoto model with a ring network topology, for which we have already shown that spurious correlations are significant in Section~\ref{sec:spurious_loc}. We will then show that our localization scheme also greatly improves DA for random Erd\H{o}s-R\'{e}nyi network topologies and scale-free Barab\'{a}si-Albert network toplogies.

In most of our analyses (all except in Section~\ref{sec:DA_KM}\ref{sec:full_sync_IC}) we initialize the Kuramoto model with a uniformly random initial condition of phases $\phi_i(0)$ on $[0,2\pi)$. We then observe the transition of the system from an initially incoherent state to a synchronized state for sufficiently large coupling strength $\kappa$. Synchronization presents a challenge for all data-driven methods. In a fully synchronized state, phases co-rotate at a uniform frequency, which is effectively a stationary state in a co-rotating reference frame. Once synchronization occurs, the observations progressively decrease in value, since they yield little new information. In Section~\ref{sec:DA_KM}\ref{sec:full_sync_IC} we discuss DA applied to cases with an initially synchronized state. We show that DA is still able to make improvements on the estimates of the phases $\phi_i$ and natural frequencies $\omega_i$, but the gain in accuracy is less than that obtained from observing the transient which carries more dynamic information.

\subsubsection{Ring network topology}

\begin{figure*}[tbp]
\centering
\includegraphics[width=\textwidth]{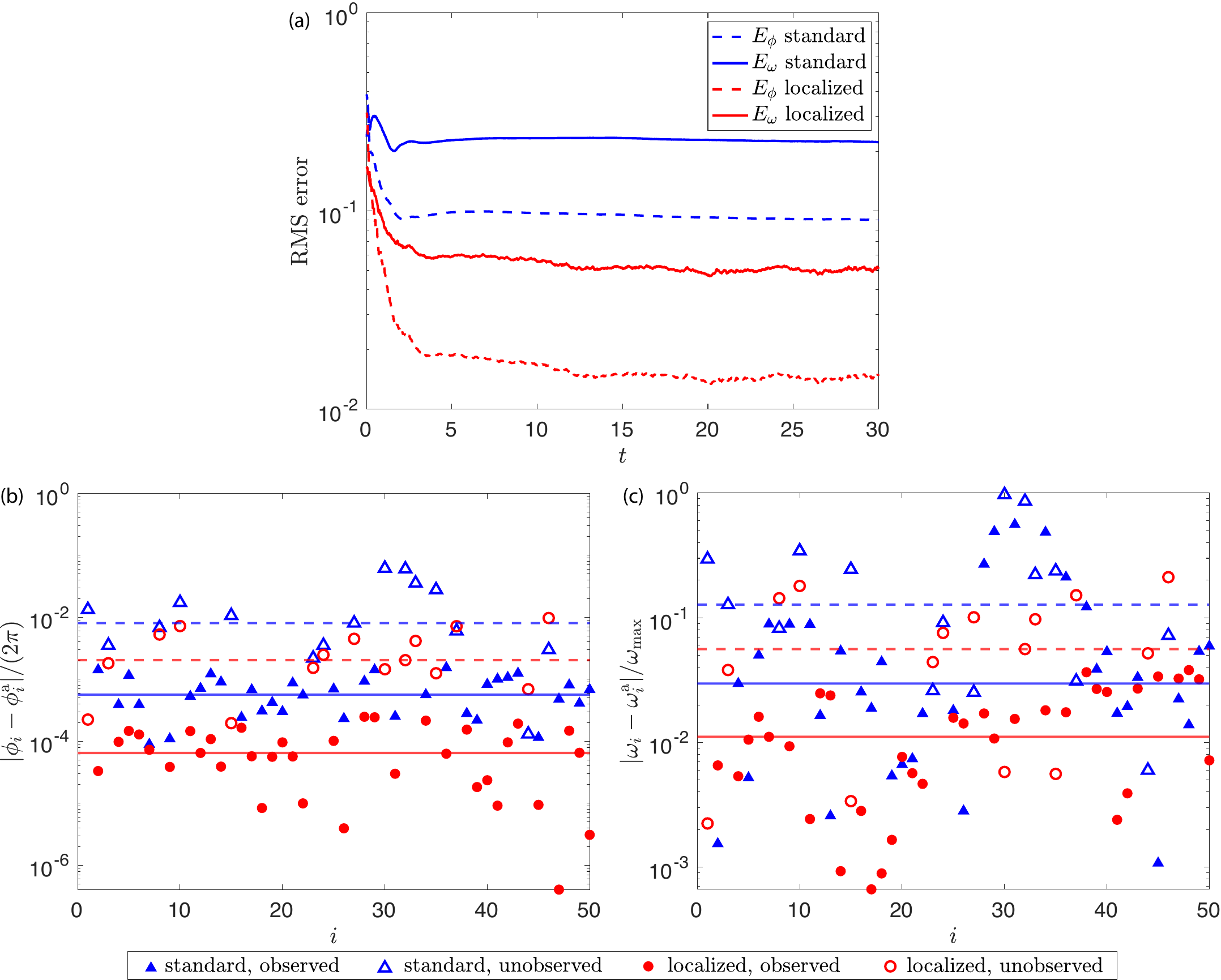}
\caption{
(a)~RMS errors in $\phi$ (dashed) and $\omega$ (solid) for DA applied to the Kuramoto model (\ref{eq:Kuramoto}) with 35 out of $N=50$ phases observed. Results are shown for the standard (blue) and localized ((\ref{eq:localization_matrix}) using $\lambda = 0.46$) (red) EnKF.  (b),~(c)~Relative errors in $\phi_i$ and $\omega_i$, respectively, at final time $t=30$ for the standard (blue triangles) and localized (red circles) EnKF. Filled symbols indicate nodes whose phase (but not frequency) is being observed, and open symbols indicate nodes whose phase and frequency are unobserved. Median errors are shown as horizontal lines; blue for standard, red for localized, solid for observed, dashed for unobserved. The Kuramoto model parameters are as in Fig.~\ref{fig:KM_phi_of_t}.
}
\label{fig:KM_DA_err}
\end{figure*}

As in Section~\ref{sec:spurious_loc}, we consider a ring network topology with $N=50$ nodes, such that the nodes are arranged on a circle, and each node is connected to its $2r$ nearest neighbors (here $r=3$).
For our DA process, we consider the case where only 35 out of 50 phases are observed. The observed set of nodes is chosen randomly. Fig.~\ref{fig:KM_DA_err}(a) shows the RMS errors between the true and the DA estimated values for the phases $\phi_i$ (dashed) and natural frequencies $\omega_i$ (solid), using both the standard (blue) and localized (red) EnKF.  Localization greatly reduces the error in both the $\phi_i$ and $\omega_i$ compared to the standard EnKF, with almost 10 times less error in each. The relative errors at time $t=30$ are shown in Fig.~\ref{fig:KM_DA_err}(b,c) for each of the $\phi_i$ and each of the $\omega_i$, respectively. It is again seen that localization (red circles) greatly reduces the error compared to the standard EnKF (blue triangles). As expected, for both methods errors are smaller for nodes whose phase is observed (filled circles/triangles) than for nodes whose phase is unobserved (open circles/triangles). Recall that none of the frequencies $\omega_i$ are observed.
As expected, errors are largest in regions with a large number of unobserved nodes, e.g., for $i\in[20,40]$. 

\begin{figure*}[tbp]
\centering
\includegraphics[width=0.5\textwidth]{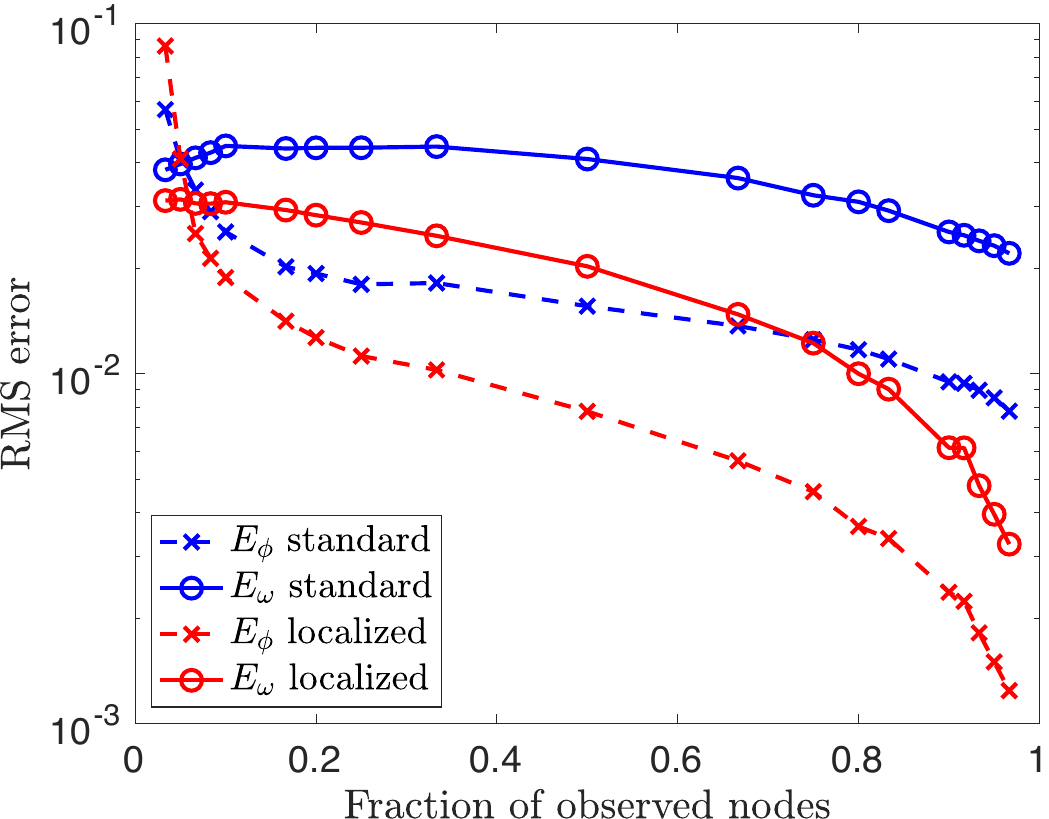}
\caption{
Median RMS errors at $t=10$ in $\phi$ (dashed) and $\omega$ (solid) for different fractions of observed phases when the EnKF is applied to the Kuramoto model (\ref{eq:Kuramoto}) with $N=60$ nodes and all other parameters the same as in Fig.~\ref{fig:KM_phi_of_t}. The median is taken over 100 random realizations of the DA process for each fraction of observed nodes. Results for the standard EnKF are shown in blue, results for the EnKF with network-specific localization are shown in red. Both DA approaches use $M=2N+1=121$ ensemble members.
}
\label{fig:KM_DA_Nobs}
\end{figure*}

As expected, the estimation accuracy increases with the fraction of observed phases. Localization amplifies this effect,  and allows for a much smaller set of observed phases to achieve the same estimation accuracy compared to the standard EnKF. We show this using a ring topology and investigate the effect of changing the number of observed phases. To simplify the analysis, we do not choose observed phases at random, but rather require that they are evenly distributed, e.g. every fourth node in the ring is observed/unobserved. For each fraction of observed phases we perform each DA method for 100 random realizations (randomized initial phases, natural frequencies, observation errors and initial ensemble) and take the median over the realizations of the RMS errors of the phases $\phi_i$ and frequencies $\omega_i$.
We consider $N=60$ which allows more data points for evenly distributed observed/unobserved nodes. The median RMS errors in $\phi$ (dashed) and $\omega$ (solid) are shown in Fig.~\ref{fig:KM_DA_Nobs} for different fractions of observed phases, for both the standard EnKF (blue) and the EnKF with network-specific localization (red). We see again a significant decrease in the RMS error when localization is employed. The trends for both the standard EnKF and localized EnKF are similar, with sharp decreases in the error of the phases $\phi_i$ for fractions of observed phases close to both 0 and 1, and with less decrease in error for fractions of observed phases close to 50\%, where there appears to be a point of inflection. For the error in the frequencies $\omega_i$, the localized EnKF shows an initially steady decrease in the error as the fraction of observed phases increases, with significant gains in accuracy for fractions of observed nodes above 50\%. 



\subsubsection{Observations from a fully synchronized state} \label{sec:full_sync_IC}

\begin{figure*}[tbp]
\centering
\includegraphics[width=0.5\textwidth]{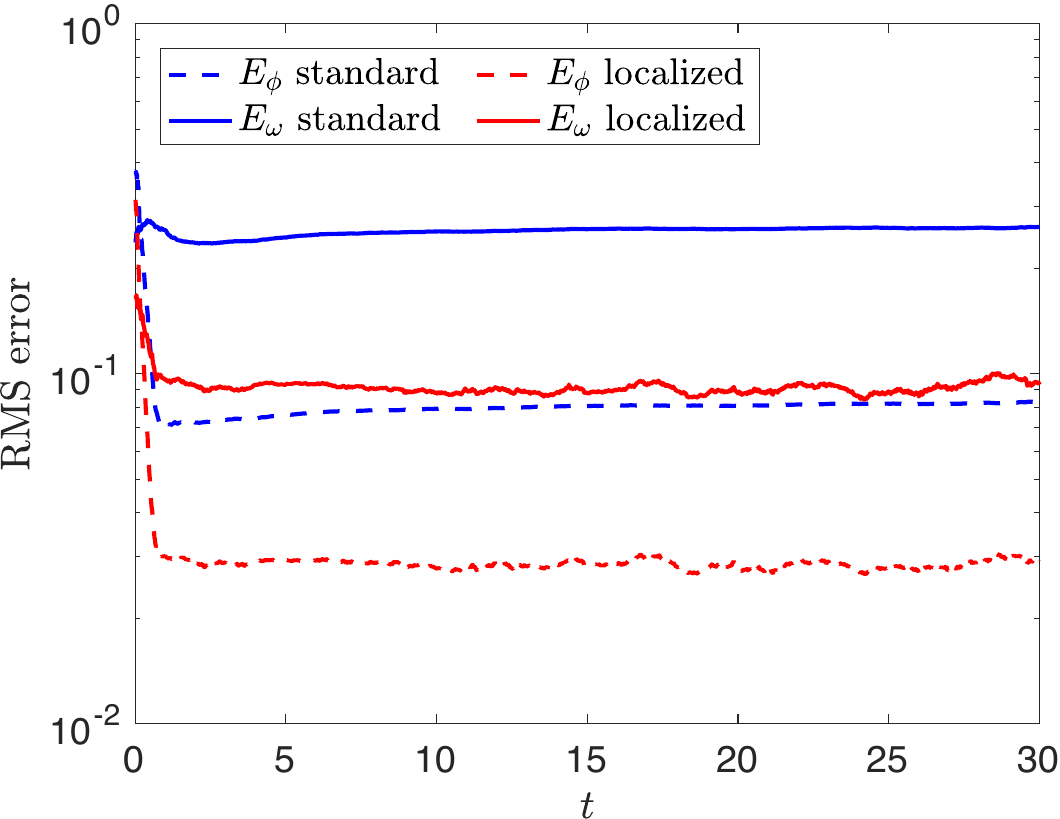}
\caption{
RMS errors in $\phi$ (dashed) and $\omega$ (solid) for DA applied to the Kuramoto model (\ref{eq:Kuramoto}) with a fully synchronized initial condition, and with 35 out of $N=50$ phases observed. Results are shown for the standard (blue) and localized ((\ref{eq:localization_matrix}) using $\lambda = 0.46$) (red) EnKF.  The Kuramoto model parameters are as in Fig.~\ref{fig:KM_phi_of_t}.
}
\label{fig:KM_DA_full_sync}
\end{figure*}

As in the previous section, we consider the Kuramoto model (\ref{eq:Kuramoto}) with a ring network topology with $r=3$, but instead of a uniformly random initial condition for the phases, we begin with a fully synchronized initial condition. Specifically, we continue the simulation shown in Fig.~\ref{fig:KM_phi_of_t} up to time $t=600$, and use the synchronized end state as the initial condition for a further DA trial. Since we have moved to a co-rotating reference frame, the synchronized state is stationary, and the observed phases are of the form $\phi_i^* + \eta_k$, where $\phi_i^*$ is the constant phase of the synchronized state and $\eta_k\sim \mathcal{N}(0,0.02^2)$ is added to create noisy observations. As is shown in Fig.~\ref{fig:KM_DA_full_sync}, DA with network specific localization is able to improve the estimate of both the phases $\phi_i$ and the frequencies $\omega_i$, but to a lesser extent than the gain that is made when the transient dynamics is observed (cf. Fig.~\ref{fig:KM_DA_err}). This is expected because a stationary state carries less information about the dynamics of the system (only an equilibrium solution). We again observe that the standard EnKF does a much worse job of estimating the phases $\phi_i$ and the frequencies $\omega_i$ compared to the localized EnKF.

\subsubsection{Random network topologies} \label{sec:DA_KM_ER}

We have used the ring topology to illustrate spurious correlations and the benefit of network-specific localization. Our methodology readily extends to general network topologies, and network-specific localization again yields great improvement in estimation accuracy. We show this for random Erd\H{o}s-R\'{e}nyi (ER) networks and random modified Barab\'{a}si-Albert (BA) networks. In ER networks, each pair of nodes is connected with probability $p$ (here we use $p=0.1$), as such ER networks are highly homogeneous, with a narrow degree distribution around the mean degree $(N-1)p$. For the BA networks, we use a modified version of the standard generative algorithm, as described in \cite{SmithGottwald21}. The algorithm starts with an initial seed network with $m_0$ nodes, and successive nodes are added iteratively. When a new node is added, $n$ new edges are added to the existing network, with preferential attachment to high degree nodes, where $n$ is drawn uniformly randomly for each new node from $\{m_1,\dots,m_2\}$, with $ m_1\geq 1$ and $ m_2 \leq m_0$. In the standard BA algorithm $m_1 = m_2 = m$. Our modified version allows for nodes with smaller degrees, including leaf nodes with degree 1 provided $m_1=1$. The BA networks are ``scale-free'', having a power law degree distribution and mean degree $m_1 + m_2$. Here we use $m_1 = 1$, $m_2 =5$, and an initial seed network with $m_0=5$ nodes.

For our network-specific localization, we use the same localization matrix $\mathcal{L}$ defined by (\ref{eq:localization_matrix_0}) and (\ref{eq:localization_matrix}) using the positive semi-definite matrix exponential $\mathcal{E} = \exp(\lambda A)$. The value of the tunable parameter $\lambda$ is determined as described in Section~\ref{sec:choosing_lambda}.

\begin{figure*}[tbp]
\centering
\includegraphics[width=\textwidth]{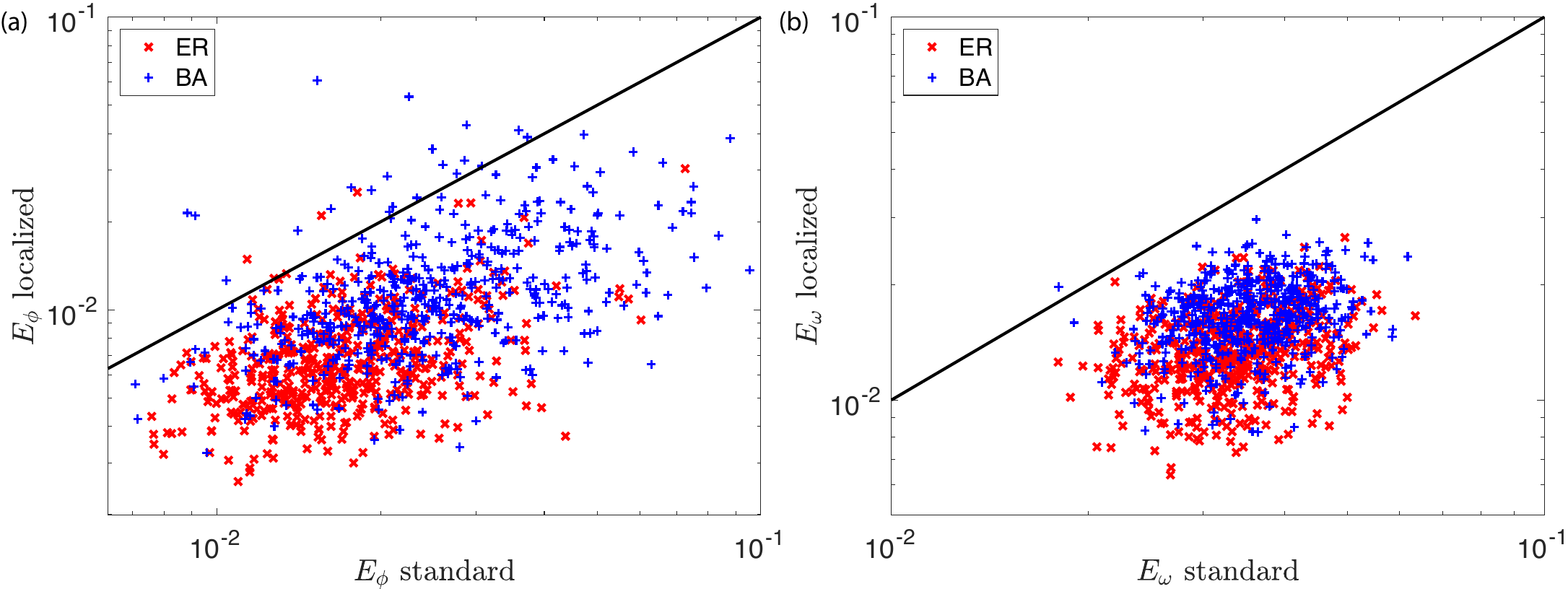}
\caption{
RMS errors at $t=10$ in (a) $\phi$, and (b) $\omega$ for DA applied to the Kuramoto model (\ref{eq:Kuramoto}) with 35 out of $N=50$ phases observed. Results are shown for 500 ER network topologies (red crosses) and 500 BA network topologies (blue pluses). All other parameters are as in Fig.~\ref{fig:KM_phi_of_t}. Errors using the standard EnKF are shown versus errors using the EnKF with network-specific localization (\ref{eq:localization_matrix_0}) and (\ref{eq:localization_matrix}) with $\lambda$ obtained from (\ref{eq:ER_lambda_2}). The black reference lines show ``standard EnKF error''=``localized EnKF error''.
}
\label{fig:KM_DA_ER}
\end{figure*}

To show the improvement that is gained by employing localization, we consider 500 realizations of the Kuramoto model with ER network topologies and 500 realizations of the Kuramoto model with BA network topologies. All realizations have $N=50$ nodes and random natural frequencies $\omega_i$ drawn from $g(\omega)\sim\mathcal{N}(0,0.1)$.  For each realization we run the EnKF with 35 out of 50 random phases observed, both with and without localization (all other parameters such as the initial ensemble are the same to allow direct comparison). We record the RMS errors in both the phases $\phi_i$ and the frequencies $\omega_i$ at time $t=10$ (which is the typical synchronization time for such a network with coupling strength $\kappa=10$, cf. Fig.~\ref{fig:KM_phi_of_t}). The results are collected in Fig.~\ref{fig:KM_DA_ER}, which shows the RMS error of the standard EnKF versus that of the localized EnKF for each of the realizations (red crosses for ER networks and black pluses for BA networks) for (a) $\phi$, and (b) $\omega$. Data points that fall below the black reference line are those for which the localized EnKF performed better (less error). Fig.~\ref{fig:KM_DA_ER}(a) shows that the localized EnKF produced a better approximation for the phases $\phi_i$ for $99.2\%$ (496/500) of the ER networks and $95.6\%$ (478/500) of the BA networks. Fig.~\ref{fig:KM_DA_ER}(b) shows that the localized EnKF produced a better prediction for the natural frequencies $\omega_i$ for every ER network, and all except one BA network ($99.8\%$). The median reduction in the error when using the localized EnKF compared to the standard EnKF is $61.0\%$ for $\phi_i$ and $59.8\%$ for $\omega_i$ in ER networks, and $53.9\%$ for $\phi_i$ and $52.8\%$ for $\omega_i$ in BA networks. It is clear that incorporating localization yields a great improvement in the estimation accuracy of DA for complex network topologies as well as the ring topology.



\subsection{Data assimilation for a network of theta neurons with a ring-like topology} \label{sec:DA_theta}

We now apply our DA approach to a network of theta neurons (\ref{eq:theta_neurons}), in which synchronization does not generally occur.

As for the Kuramoto model, our task is to estimate all the phases $\phi_i$ and all the intrinsic firing parameters $\zeta_i$ when only a subset of the phases are observed.

For the connectivity matrix $B$, we consider a ring topology, such that nodes are positively coupled with $B_{ij}=1$ if they are within a coupling radius $r=3$. We also include negative (inhibitory) long-range coupling, such that $B_{ij}=-0.4$ for the three furthest nodes from each node. All other entries of $B$ are zero. This connectivity matrix produces the ``bump state'' shown in Fig.~\ref{fig:theta_DA_1}.
Similar to the Kuramoto model, correlation matrices for large ensembles reflect the underlying network connectivity. The localization matrix (\ref{eq:localization_matrix}) is chosen, with $A_{ij} = |B_{ij}|$ since $B$ has negative entries. We use $\lambda=0.46$, as determined in Section~\ref{sec:choosing_lambda} for a ring topology with coupling radius $r=3$.

\begin{figure*}[tbp]
\centering
\includegraphics[width=\textwidth]{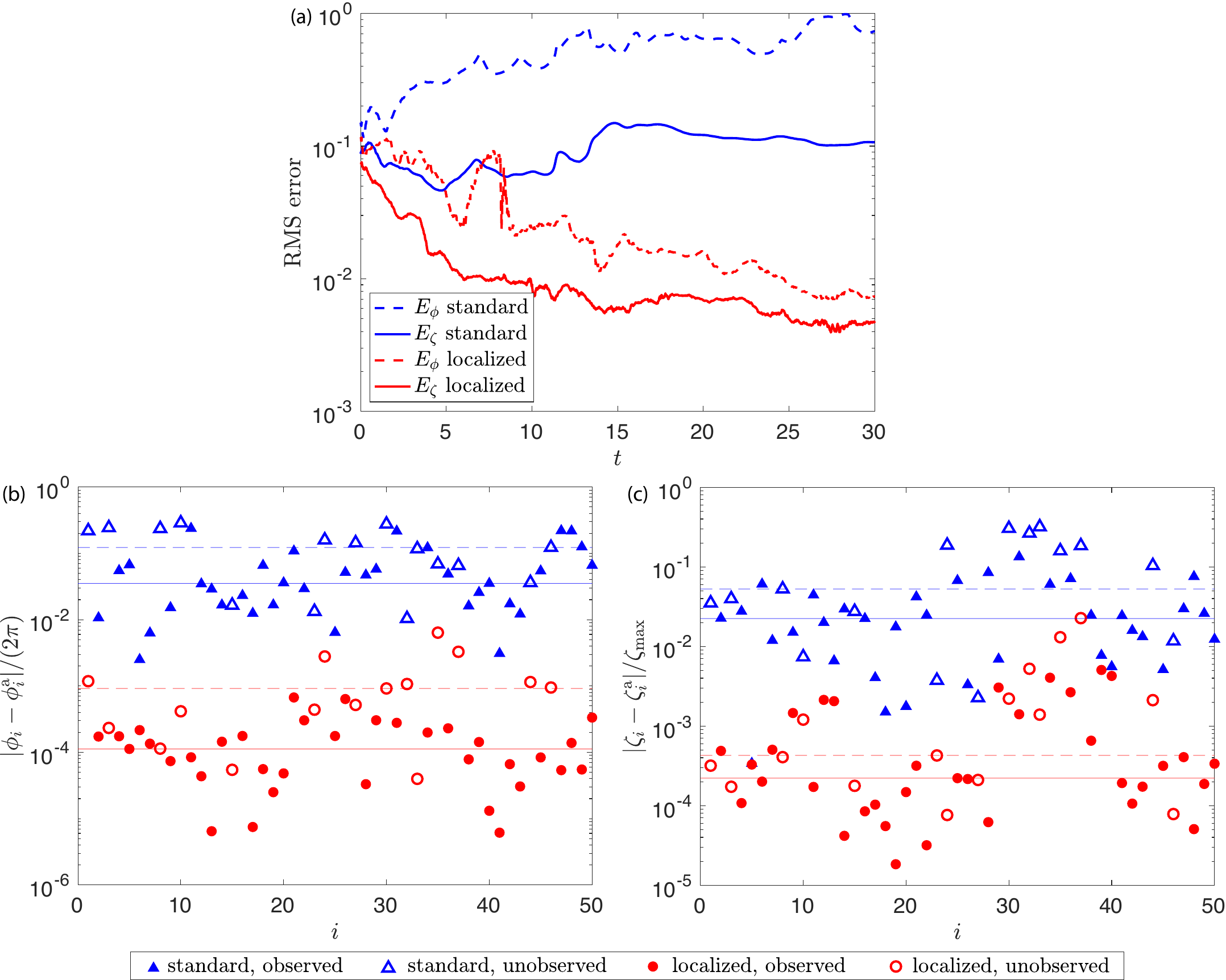}
\caption{
(a)~RMS errors in $\phi$ (dashed) and $\zeta$ (solid) for DA applied to a theta neuron network (\ref{eq:theta_neurons}) with 35 out of $N=50$ phases observed. Results are shown for the standard (blue) and localized ((\ref{eq:localization_matrix}) using $\lambda=0.46$) (red) EnKF. (b),~(c)~Relative errors in $\phi_i$ and $\zeta_i$, respectively, at final time $t=30$ for the standard (blue triangles) and localized (red circles) EnKF. Filled symbols indicate observed nodes, and open symbols indicate unobserved nodes. Median errors are shown as horizontal lines; blue for standard, red for localized, solid for observed, dashed for unobserved. The theta neuron model parameters are as in Fig.~\ref{fig:theta_DA_1}. 
}
\label{fig:theta_DA_err}
\end{figure*}

We find that localization again significantly improves the estimation capability of the EnKF, and yields accurate estimates for both the $\phi_i$ and $\zeta_i$ when only 35 out of 50 phases are observed. The RMS errors in $\phi_i$ (dashed) and $\zeta_i$ (solid) are shown in Fig.~\ref{fig:theta_DA_err}(a) for the standard (blue) and localized (red) EnKF. By time $t=30$, localization has reduced the RMS errors in both the $\phi_i$ and $\zeta_i$ by more than a factor of 10. The relative errors at time $t=30$ are shown in Fig.~\ref{fig:theta_DA_err}(b,c). The errors clearly show excellent performance of the EnKF approach with localization. Comparing the median errors (horizontal lines in Fig.~\ref{fig:theta_DA_err}(b,c)), localization yields more than a 100-fold decrease in error compared to the standard EnKF. As with the Kuramoto model, there is greater discrepancy for the unobserved nodes (open circles/triangles) compared to the observed nodes (filled circles/triangles). For the quiescent, approximately stationary, nodes in the bump state ($i\in[30,40]$), DA performs slightly worse. This is analogous to the problem of redundant observations in synchronization, though the persistent small fluctuations that occur in the theta neuron model allow for better approximation than a purely stationary synchronized state.

\begin{figure*}[tbp]
\centering
\includegraphics[width=0.5\textwidth]{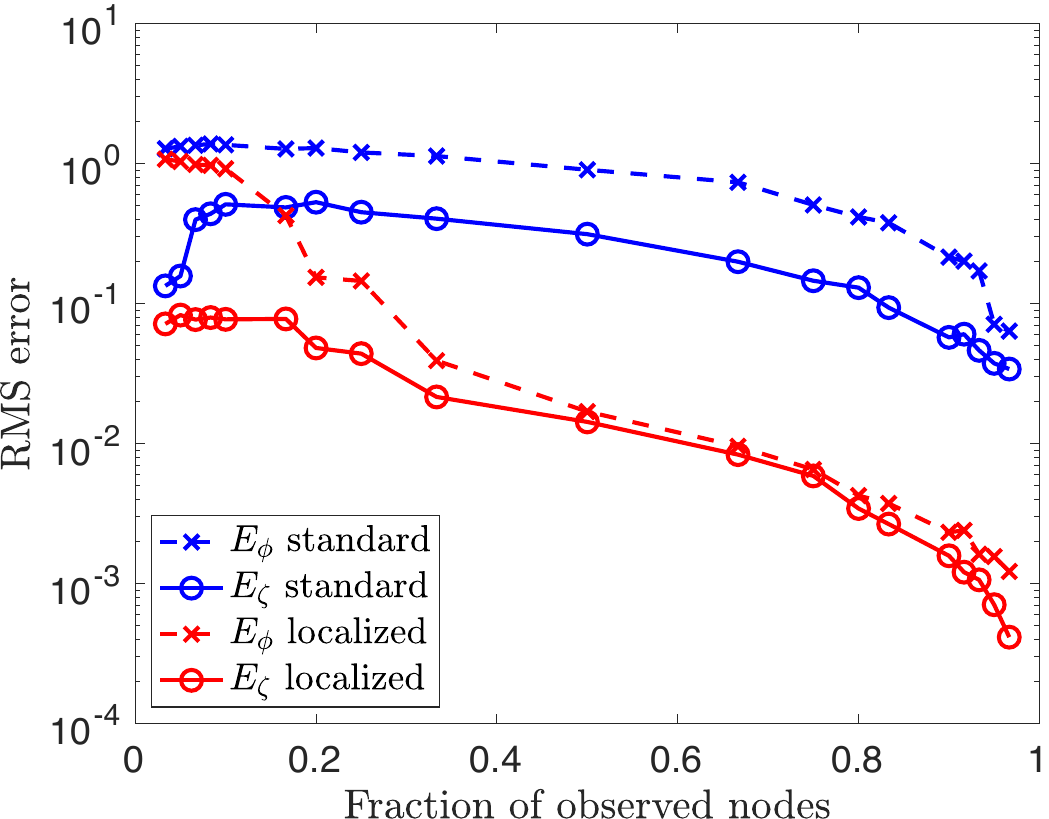}
\caption{
Median RMS errors at $t=30$ in $\phi$ (dashed) and $\zeta$ (solid) for different fractions of observed phases when the EnKF is applied to the theta neuron model (\ref{eq:theta_neurons}) with $N=60$ nodes and all other parameters the same as in Fig.~\ref{fig:theta_DA_1}. The median is taken over 100 random realizations of the DA process for each fraction of observed phases. Results for the standard EnKF are shown in blue, results for the EnKF with network-specific localization are shown in red. Both DA approaches use $M=2N+1=121$ ensemble members.
}
\label{fig:theta_DA_Nobs}
\end{figure*}

Fig.~\ref{fig:theta_DA_Nobs} shows that, as expected, when the fraction of observed phases increases, the estimation accuracy of both the standard and localized EnKF increases. However, the EnKF with localization allows for a much smaller number of observed phases to achieve the same estimation accuracy compared to the standard EnKF. For example, to achieve the same RMS errors as the standard EnKF with 90\% of the phases observed, one only needs to observe 20\% of the phases using the localized EnKF.

DA performs better with overall smaller RMS errors for the theta neuron model (\ref{eq:theta_neurons}) than for the Kuramoto model (\ref{eq:Kuramoto}), because its dynamics is irregular, and, thus, DA does not suffer from the data degeneracy issue that arises due to synchronization.

\section{Conclusion} \label{sec:conclusions}
Data assimilation via the Ensemble Kalman Filter with network-specific localization and state space augmentation (and other minor modifications to account for periodicity of phase variables) can accurately determine both the phases and unknown model parameters in networks of coupled oscillators when only a subset of the phases are observed. Our novel localization approach utilizes the matrix exponential of the network's adjacency matrix, which encodes the connectivity between nodes. We have demonstrated the efficacy of our method for the Kuramoto model that exhibits synchronization, and networks of theta neurons which do not exhibit synchronization. In both examples, data assimilation yields excellent approximations that closely agree with the truth. 

We have considered random subsets for the observed nodes, but it is likely that there are optimal sets of nodes that should be observed. Real-world applications such as the power grid will also have physical limitations on which nodes can be observed. More work is needed to determine an optimal choice of observed nodes.

We have focused on using DA to learn the unknown intrinsic parameters for each oscillator. Future work should investigate whether DA with partial observations can be used to determine unknown coupling functions (estimating Fourier coefficients) or unknown network connectivity matrices\cite{Peixoto18, Peixoto19, TylooEtAl2021, DengEtAl2022, ChenEtAl2022, GaskinEtAl2023}. Determining the network structure is challenging since there is a large number of unknowns, and our localization method requires an \textit{a priori} known network structure. However, we believe that if the network topology is known reasonably well, e.g., it is known where power lines are in the power grid, but the non-zero weights of the adjacency matrix are not known, then localization can still be performed and the filter will converge. 

We have shown the effectiveness of using the matrix exponential of the network adjacency matrix as a localization matrix for dynamics on networks, but future work should consider other types of localization matrices for networks. For instance, sparse localization matrices could be considered, which improve computational efficiency (see Section~\ref{sec:Gaspari-Cohn}), or transform-then-localize schemes could be employed \cite{SnyderHakim22}. 


In recent years there has been a surge in data-driven methods to perform inverse problems, such as entropic outlier sparsification and entropy-optimal scalable probabilistic approximation \cite{Horenko22, HorenkoEtAl23, GroomEtAl24}. It would be interesting to see if our localization method can be beneficial for these methods as well.

Our novel localization method is specific to dynamics on networks, but not necessarily to coupled oscillators. We expect our localization approach to also improve DA when applied to other types of dynamics on networks, such as spreading processes (e.g., contagions) \cite{DeDomenicoEtAl2016}.

\enlargethispage{20pt}

\vspace{12pt}

\noindent \textbf{Acknowledgements}

\noindent We wish to acknowledge support from the Australian Research Council, Grant No. DP180101991, and from the Marsden Fund of the Royal Society of New Zealand, managed by Royal Society Te Apārangi, Project ID 23-UOA-152.

\vspace{12pt}

\noindent \textbf{Data availability}

\noindent The code that implements our methodology, generates all data used, and processes the data to create the figures in the manuscript can be found at Zenodo: \url{https://doi.org/10.5281/zenodo.15092531}


\vskip2pc

%

\end{document}